\documentclass[10pt,aps,twocolumn,prc,showpacs,nofootinbib,noshowkeys,superscriptaddress,floatfix]{revtex4-1}

\usepackage{amsmath,amsfonts,amsthm,amssymb}
\usepackage[dvips]{graphics,graphicx}
\usepackage[usenames,dvipsnames]{color}
\definecolor{darkblue}{RGB}{0,0,196}
\definecolor{darkgreen}{RGB}{0,120,0}
\usepackage[colorlinks=true,linktocpage=true,linkcolor=darkblue,citecolor=red,urlcolor=darkblue]{hyperref}

\usepackage{cancel}
\usepackage{multirow}
\usepackage{longtable}
\usepackage{color}

\begin{document}

\preprint{}

\title{Exact solutions and attractors of higher-order viscous fluid dynamics for Bjorken flow}

\author{Sunil Jaiswal}
\email{sunil.jaiswal@tifr.res.in}
\affiliation{Department of Nuclear and Atomic Physics, Tata Institute of Fundamental Research, Mumbai 400005, India}
\author{Chandrodoy Chattopadhyay}
\email{chattopadhyay.31@osu.edu}
\affiliation{Department of Physics, The Ohio State University, Columbus, Ohio 43210-1117, USA}
\author{Amaresh Jaiswal}
\email{a.jaiswal@niser.ac.in}
\affiliation{School of Physical Sciences, National Institute of Science Education and Research, HBNI, Jatni 752050, Odisha, India}
\author{Subrata Pal}
\email{spal@tifr.res.in}
\affiliation{Department of Nuclear and Atomic Physics, Tata Institute of Fundamental Research, Mumbai 400005, India}
\author{Ulrich Heinz}
\email{heinz.9@osu.edu}
\affiliation{Department of Physics, The Ohio State University, Columbus, Ohio 43210-1117, USA}
\affiliation{Institut f\"ur Theoretische Physik, J.\,W.\,Goethe-Universit\"at, Max-von-Laue-Stra\ss e\,1, D-60438 Frankfurt am Main, Germany}

\date{\today}

\begin{abstract}
We consider causal higher order theories of relativistic viscous hydrodynamics in the limit of one-dimensional boost-invariant expansion and study the associated dynamical attractor. We obtain evolution equations for the inverse Reynolds number as a function of Knudsen number. The solutions of these equations exhibit attractor behavior which we analyze in terms of Lyapunov exponents using several different techniques. We compare the attractors of the second-order M\"uller-Israel-Stewart (MIS), transient Denicol-Niemi-Molnar-Rischke (DNMR), and third-order theories with the exact solution of the Boltzmann equation in the relaxation-time approximation. It is shown that for Bjorken flow the third-order theory provides a better approximation to the exact kinetic theory attractor than both MIS and DNMR theories. For three different choices of the time dependence of the shear relaxation rate we find analytical solutions for the energy density and shear stress and use these to study the attractors analytically. By studying these analytical solutions at both small and large Knudsen numbers we characterize and uniquely determine the attractors and Lyapunov exponents. While for small Knudsen numbers the approach to the attractor is exponential, strong power-law decay of deviations from the attractor and rapid loss of initial state memory is found even for large Knudsen numbers. Implications for the applicability of hydrodynamics in far-off-equilibrium situations are discussed.
\end{abstract}

\pacs{25.75.-q, 24.10.Nz, 47.75+f}


\maketitle

\section{Introduction}
\label{intro}
\vspace*{-2mm}

Hydrodynamics is an effective macroscopic theory describing the large scale and slowly varying dynamical modes of multiparticle systems. The conventional formulation of hydrodynamic equations proceeds by assuming a separation of macroscopic and microscopic length and time scales such that gradients of local-equilibrium quantities, normalized to appropriate powers of the system's energy density, are small and one can perform an order-by-order gradient expansion around local thermodynamic equilibrium \cite{Florkowski:2017olj}. Therefore the remarkable success of relativistic hydrodynamics in describing the quark-gluon plasma (QGP) formed in ultra-relativistic heavy-ion collisions initially led to the belief that these collisions create a nearly thermalized medium close to local thermal equilibrium \cite{Heinz:2001xi} (see also the reviews \cite{Heinz:2013th, Jaiswal:2016hex, Florkowski:2017olj, Romatschke:2017ejr}). On the other hand, with the advent of numerical dissipative relativistic fluid dynamics \cite{Heinz:2005bw, Romatschke:2007mq, Song:2007fn, Dusling:2007gi, Song:2007ux, Luzum:2008cw, Song:2008si} it became clear that the dynamical evolution of heavy-ion collisions is affected by persistent large dissipative corrections. In spite of this numerical evidence for large deviations from local thermal equilibrium, second-order dissipative relativistic fluid dynamics (especially when coupled with a hadronic cascade to describe the final dilute decoupling stage \cite{Song:2010aq, Heinz:2011kt}) met with impressive phenomenological success and predictive power in the description of heavy-ion collision experiments \cite{Romatschke:2007mq, Luzum:2008cw, Song:2008si, Song:2008hj, Schenke:2010nt, Song:2010mg, Song:2011hk, Song:2011qa, Shen:2011eg, Schenke:2011bn, Gale:2012rq}, and was even successfully applied to explain flow data in small collision systems formed in proton-proton (p-p) and proton-lead (p-Pb) collisions \cite{Bozek:2015swa, Bozek:2016jhf}. This ``unreasonable effectiveness'' of hydrodynamics as a dynamical description of high-energy hadronic collisions in situations that are even very far away from local thermal equilibrium has generated much recent interest in the very foundations of fluid dynamics \cite{Heller:2013fn, Heller:2015dha, Aniceto:2015mto, Basar:2015ava, Florkowski:2016zsi, Heller:2016rtz, Romatschke:2016hle, Behtash:2017wqg, Blaizot:2017lht, Blaizot:2017ucy, Denicol:2017lxn, Romatschke:2017acs, Romatschke:2017vte, Strickland:2017kux, Spalinski:2017mel, Aniceto:2018uik, Behtash:2018moe, Chattopadhyay:2018apf, Denicol:2018pak, Heller:2018qvh, Strickland:2018ayk, Tinti:2018qfb, Behtash:2019txb, Blaizot:2019scw, Grozdanov:2019uhi, Grozdanov:2019kge, Heinz:2019dbd, Strickland:2019hff}, culminating in the formulation of a new ``far-from-local-equilibrium fluid dynamics'' paradigm \cite{Romatschke:2017ejr, Romatschke:2017vte}. The present work is a contribution to this ongoing discussion, adding new analytic results for the heavily studied simple case of (0+1)-dimensional Bjorken expansion of a transversally homogeneous system with longitudinal boost-invariance \cite{Bjorken:1982qr}. 

The simplest relativistic dissipative theory, relativistic Navier-Stokes theory \cite{Eckart:1940zz, Landau:1987}, imposes instantaneous constitutive relations between the dissipative flows and their generating forces, expressed through first-order gradients of equilibrium quantities. This approach was found to be plagued by  acausality and intrinsic instability \cite{Hiscock:1983zz, Hiscock:1985zz}. The phenomenological second-order theory developed by M\"uller, Israel and Stewart (MIS) \cite{Muller:1967zza, Israel:1976tn, Israel:1979wp} cures these problems by introducing a relaxation type equation for the dissipative flows and thus turning them into independent dynamical degrees of freedom of the system whose evolution is controlled by the competition between macroscopic expansion (driving the system away from local equilibrium) and microscopic scattering (driving it back towards local equilibrium). As discussed in \cite{Romatschke:2017ejr}, even the minimal causal conformal theory given by MIS introduces new modes called non-hydrodynamic modes that were absent in Navier-Stokes theory. These non-hydrodynamic modes are now known to play an important role in the approach to the regime of applicability of hydrodynamics, also known as the ``hydrodynamization'' process \cite{Heller:2014wfa, Heller:2015dha, Kurkela:2015qoa, Heller:2016rtz, Heller:2016gbp}. For conformal systems undergoing longitudinal Bjorken expansion, it was shown explicitly that hydrodynamization occurs at microscopic thermalization time scales which, in strongly and anisotropically expanding systems, are much shorter than both the time scales for local isotropization and thermalization \cite{Romatschke:2017vte}. Similar conclusions were obtained in recent studies \cite{Behtash:2017wqg, Chattopadhyay:2018apf} of systems undergoing Gubser flow \cite{Gubser:2010ze} where the fireball expands not only boost-invariantly in longitudinal, but also azimuthally symmetrically in transverse direction. These findings lead to the conclusion that the criterium of proximity to local thermodynamic equilibrium for hydrodynamics to be valid is (at least in these carefully analytically studied simplified situations) too strict and should be replaced by a different condition stating that contributions from the non-hydrodynamic modes can be neglected \cite{Romatschke:2016hle}. In the present study, we will focus on yet another interesting feature that appears in a causal theory of relativistic dissipative hydrodynamics, ``the hydrodynamic attractor" \cite{Heller:2015dha,Romatschke:2017acs,Denicol:2018pak,Behtash:2017wqg,Strickland:2017kux,Chattopadhyay:2018apf}, and its role in the hydrodynamization process.

Attractor behavior was first identified by considering the hydrodynamic formulation as a gradient series expansion. Recently, this gradient series was shown, for several highly symmetric flow configurations that are amenable to analytic treatment, to have zero radius of convergence. This suggests that hydrodynamic theory cannot be systematically improved by taking into account higher order-terms in the gradient series \cite{Heller:2015dha}. The gradient expansion generates an asymptotic series which exhibits initial signs of convergence for a few terms before eventually diverging \cite{Heller:2015dha, Basar:2015ava}. The initial appearance of convergence may explain the observed remarkable phenomenological success of hydrodynamic formulations based on truncations of the gradient expansion at second or third order but, due to the ultimate divergence of the series, the theory cannot be improved beyond a certain order by keeping additional terms. Fortunately, the (diverging) gradient expansion series can be Borel resummed, giving rise to a unique hydrodynamic attractor solution (hydrodynamic mode), {\it which is well defined even for large gradients}, and a series of rapidly decaying non-hydrodynamic modes that describe the approach towards this attractor from arbitrary initial conditions \cite{Heller:2013fn, Heller:2015dha}. This suggests that hydrodynamics displays a novel type of universality even far from local equilibrium which is independent of the initial state of the system, indicating the existence of a new, far-from-local-equilibrium hydrodynamic theory \cite{Romatschke:2017vte}. Recently, it was discovered \cite{Strickland:2019jut} that this phenomenon extends even beyond hydrodynamics: in kinetic theory also the evolution of non-hydrodynamic higher-order momentum moments of the distribution function is controlled by attractors. 

The theory of dynamical attractors can be complex, and hence it is instructive to have examples were the attractor is quantitatively understood using analytical methods. Perhaps the most useful quantities associated with an attractor are Lyapunov exponents which characterize the rate of separation of infinitesimally close trajectories in the phase-space evolution of dynamical systems \cite{Goldstein}. While negative Lyapunov exponents are associated with dissipative systems and indicate the existence of attractors, positive values are usually associated with chaotic systems. For a conservative system one obtains vanishing Lyapunov exponents.  In this article, we employ the theory of Lyapunov exponents to study the attractors for MIS theory and two other, improved versions of causal relativistic dissipative hydrodynamics (see Refs.~\cite{Behtash:2017wqg, Behtash:2019txb} for earlier related work).

Motivated by its application to QGP evolution, several improvements over MIS theory were proposed during the last decade \cite{Martinez:2010sc, Florkowski:2010cf, Denicol:2012cn, Strickland:2014pga, Bazow:2013ifa,  Jaiswal:2013vta, Jaiswal:2014isa, Tinti:2015xwa, Molnar:2016vvu, Molnar:2016gwq, McNelis:2018jho, Nopoush:2019vqc}. For Bjorken and Gubser flows, some of these theories lead to very good agreement with the exact solution of the Boltzmann equation even in situations where the deviations from thermal equilibrium are large \cite{Florkowski:2013lya, Martinez:2017ibh, Jaiswal:2013vta, Chattopadhyay:2018apf}. In this article, we consider the Denicol-Niemi-Molnar-Rischke (DNMR) theory \cite{Denicol:2012cn} which is an improved version of second-order MIS theory,  as well as a third-order hydrodynamic theory derived form relativistic kinetic theory by going to third order in the Chapman-Enskog expansion \cite{Jaiswal:2013vta}. To understand the emergent attractor behavior we go beyond previous numerical studies of these theories by finding analytical solutions of the  corresponding hydrodynamic evolution equations \cite{Chattopadhyay:2018pwe}. Analytical solutions of higher-order dissipative hydrodynamics were found previously for a few very special cases \cite{Hatta:2014gqa, Denicol:2017lxn}. We here expand this portfolio and use the new analytical solutions to study their dependence on initial conditions, their attractors, and their late-time behavior.

The rest of this article is structured as follows. In Sec.~\ref{hydro} we briefly review causal theories of second-order (MIS and DNMR) and third-order relativistic viscous hydrodynamics. In Sec.~\ref{bjorken} we simplify these theories for Bjorken flow and obtain a generic evolution equation for the inverse Reynolds number as a function of the inverse Knudsen number, which can be adapted to all three hydrodynamic theories by adjusting a set of two parameters. This ordinary differential equation is decoupled from the hydrodynamic evolution of the energy density (but feeds back into it), and thus it can be solved independently. In Sec.~\ref{Lyapunov} we demonstrate numerically that the solutions of these differential equations exhibit attractor-like behavior which we analyze in terms of Lyapunov exponents, using several different approaches. We also compare the exact numerical attractors of  the second-order MIS and DNMR theories, as well as our third-order theory, with the exact solution of the relativistic Boltzmann equation in the relaxation-time approximation (RTA). We show that the third-order theory provides a better approximation to the exact kinetic theory attractor than both the MIS and DNMR attractors. In Sec.~\ref{analy_sol} we provide further clarification of the numerical behavior found in the preceding sections by working out explicit analytical solutions for the energy density and shear stress as a function of the inverse Knudsen number, for three different forms of the shear relaxation time. These analytical solutions allow us to also study the attractors analytically. In Sec.~\ref{analy_attr}, finally, we explore the universal behavior of these solutions at both small and large Knudsen numbers and use the results to characterize and  uniquely determine the hydrodynamic attractor and its associated Lyapunov exponents in a new way. Our results are summarized and some conclusions offered in Sec.~\ref{conclusion}.


\vspace*{-2mm}
\section{Relativistic dissipative hydrodynamics}
\label{hydro}
\vspace*{-2mm}

In this section, we will briefly review the equations for relativistic dissipative hydrodynamics. We consider a conformal system, which corresponds in kinetic theory to a system of massless particles. The energy-momentum tensor for such a system, in the Landau frame, has the form
\begin{align}\label{tmunu}
T^{\mu\nu} = \epsilon u^\mu u^\nu - P\Delta ^{\mu \nu} +  \pi^{\mu\nu},  
\end{align}
where $\epsilon$ and $P$ are the local energy density and pressure. Conformal symmetry implies an equation of state (EoS) $\epsilon=3P$ and zero bulk viscous pressure, $\Pi=0$. We also define $\Delta^{\mu\nu} \equiv g^{\mu\nu}{-}u^{\mu}u^{\nu}$ which serves as a projection operator to the space orthogonal to $u^{\mu}$ (i.e. onto the spatial directions in the local rest frame (LRF)). The shear stress tensor, $\pi^{\mu\nu}$, is traceless and orthogonal to $u^{\mu}$. The metric convention used here is $g^{\mu\nu} = {\rm diag}(+\,-\,-\,-)$.

Evolution equations for $\epsilon$ and $u^\mu$ are obtained from energy-momentum conservation, $D_\mu T^{\mu\nu}=0$:
\begin{align}
\dot\epsilon + (\epsilon+P)\theta - \pi^{\mu\nu}\sigma_{\mu\nu} &= 0, \label{evol01}\\
(\epsilon{+}P)\,\dot u^\alpha - \nabla^\alpha P + \Delta^\alpha_\nu \partial_\mu \pi^{\mu\nu}  &= 0. \label{evol02}
\end{align}
Here we use the notations $D_\mu$ for the covariant derivative, $\dot A\equiv u^\mu D_\mu A$ for the co-moving time derivative, $\nabla^\alpha\equiv\Delta^{\mu\alpha} D_\mu$ for space-like derivative, $\theta\equiv D_\mu u^\mu$ for the expansion scalar, and $\sigma_{\mu\nu}\equiv \frac{1}{2} (\nabla_{\mu}u_{\nu}{+}\nabla_{\nu}u_{\mu}) - \frac{1}{3} \theta \Delta_{\mu\nu}$ for the velocity shear tensor.

To close Eqs.~(\ref{evol01}),(\ref{evol02}) we need additional equations for the shear stress $\pi^{\mu\nu}$. The simplest form of $\pi^{\mu\nu}$ is the Navier-Stokes form, which is first order in velocity gradients, $\pi^{\mu\nu}_\mathrm{NS} = 2\eta\sigma^{\mu\nu}$, where $\eta$ is the shear viscosity coefficient.  As already mentioned in the Introduction, relativistic Navier-Stokes theory violates causality and is unstable. The simplest way to restore causality is by introducing a relaxation-type equation for $\pi^{\mu \nu}$. This prescription, also known as the ``Maxwell-Cattaneo law", requires that the dissipative forces relax to their Navier-Stokes values in some finite relaxation time, i.e., $\tau_\pi\dot\pi^{\langle\mu\nu\rangle} + \pi^{\mu\nu} = 2\eta\sigma^{\mu\nu}$,\footnote{%
	Angular brackets around pairs of Lorentz indices indicate projection of the tensor onto its traceless and locally spatial part, e.g., 
	$\dot{\pi}^{\langle\mu\nu\rangle} = \Delta^{\mu\nu}_{\alpha\beta} \dot{\pi}^{\alpha\beta}$, 
	where $\Delta^{\mu\nu}_{\alpha\beta}=\frac{1}{2}(\Delta^\mu_\alpha\Delta^\nu_\beta{+}\Delta^\mu_\beta
	\Delta^\nu_\alpha) - \frac{1}{3}\Delta^{\mu\nu}\Delta_{\alpha\beta}$.}
where $\tau_{\pi}$ is the shear relaxation time. For conformally symmetric systems one more term must be added in the evolution of shear stress \cite{Baier:2006um, Baier:2007ix}:
\begin{equation}\label{MIS}
   \tau_\pi\dot\pi^{\langle\mu\nu\rangle} + \pi^{\mu\nu} 
   = 2\eta\sigma^{\mu\nu} -\frac{4}{3}\tau_\pi\pi^{\mu\nu}\theta.
\end{equation}
This equation is a close variant \cite{Baier:2006um} of the one first derived by M\"uller, Israel and Stewart \cite{Israel:1979wp, Muller:1967zza, Israel:1976tn}, and we will therefore refer to it as the ``MIS'' theory. Its derivation was based on an analysis of the entropy current and the second law of thermodynamics, without recourse to a specific theory for the underlying microscopic dynamics.\footnote{\label{fnMIS}%
	It should be noted that Eq.~(\ref{MIS}) does not include all possible second-order terms 
	\cite{Baier:2007ix}, and that (unlike the theories discussed below) its transport coefficients are 
	not matched to an underlying kinetic theory.}
 
A systematic derivation of second-order (``transient'') relativistic fluid dynamics from relativistic kinetic theory, using an expansion of the dissipative flows in momentum-moments of the distribution function, was performed in \cite{Denicol:2012cn}. For conformal systems and an RTA collision term, the result obtained in the 14-moment approximation differs from Eq.~(\ref{MIS}) by two additional terms that are of second order in gradients:
\begin{equation}
\label{DNMR}
  \dot{\pi}^{\langle\mu\nu\rangle} \!+ \frac{\pi^{\mu\nu}}{\tau_\pi}\!= 
  2\beta_{\pi}\sigma^{\mu\nu}
  \!+2\pi_\gamma^{\langle\mu}\omega^{\nu\rangle\gamma}
  \!-\frac{10}{7}\pi_\gamma^{\langle\mu}\sigma^{\nu\rangle\gamma} 
  \!-\frac{4}{3}\pi^{\mu\nu}\theta.
\end{equation}
Here $\beta_\pi\equiv\eta/\tau_\pi = 4P/5$, while  $\omega^{\mu\nu}\equiv\frac{1}{2}(\nabla^\mu u^\nu{-}\nabla^\nu u^\mu)$ is the vorticity tensor. This ``DNMR'' theory \cite{Denicol:2012cn} can also be derived from a Chapman-Enskog like iterative solution of the RTA Boltzmann equation \cite{Jaiswal:2013npa}. 

Carrying the Chapman-Enskog expansion to one additional order, a third-order evolution equation for the shear stress was derived for the same system in \cite{Jaiswal:2013vta}:
\begin{align}\label{TOSHEAR}
\dot{\pi}^{\langle\mu\nu\rangle} =& -\frac{\pi^{\mu\nu}}{\tau_\pi}
+2\beta_\pi\sigma^{\mu\nu}
+2\pi_{\gamma}^{\langle\mu}\omega^{\nu\rangle\gamma}
-\frac{10}{7}\pi_\gamma^{\langle\mu}\sigma^{\nu\rangle\gamma}  \nonumber \\
&-\frac{4}{3}\pi^{\mu\nu}\theta
+\frac{25}{7\beta_\pi}\pi^{\rho\langle\mu}\omega^{\nu\rangle\gamma}\pi_{\rho\gamma}
-\frac{1}{3\beta_\pi}\pi_\gamma^{\langle\mu}\pi^{\nu\rangle\gamma}\theta \nonumber \\
&-\frac{38}{245\beta_\pi}\pi^{\mu\nu}\pi^{\rho\gamma}\sigma_{\rho\gamma}
-\frac{22}{49\beta_\pi}\pi^{\rho\langle\mu}\pi^{\nu\rangle\gamma}\sigma_{\rho\gamma} \nonumber \\
&-\frac{24}{35}\nabla^{\langle\mu}\left(\pi^{\nu\rangle\gamma}\dot u_\gamma\tau_\pi\right)
+\frac{4}{35}\nabla^{\langle\mu}\left(\tau_\pi\nabla_\gamma\pi^{\nu\rangle\gamma}\right) \nonumber \\
&-\frac{2}{7}\nabla_{\gamma}\left(\tau_\pi\nabla^{\langle\mu}\pi^{\nu\rangle\gamma}\right)
+\frac{12}{7}\nabla_{\gamma}\left(\tau_\pi\dot u^{\langle\mu}\pi^{\nu\rangle\gamma}\right) \nonumber \\
&-\frac{1}{7}\nabla_{\gamma}\left(\tau_\pi\nabla^{\gamma}\pi^{\langle\mu\nu\rangle}\right)
+\frac{6}{7}\nabla_{\gamma}\left(\tau_\pi\dot u^{\gamma}\pi^{\langle\mu\nu\rangle}\right) \nonumber \\
&-\frac{2}{7}\tau_\pi\omega^{\rho\langle\mu}\omega^{\nu\rangle\gamma}\pi_{\rho\gamma}
-\frac{2}{7}\tau_\pi\pi^{\rho\langle\mu}\omega^{\nu\rangle\gamma}\omega_{\rho\gamma} \nonumber \\
&-\frac{10}{63}\tau_\pi\pi^{\mu\nu}\theta^2
+\frac{26}{21}\tau_\pi\pi_\gamma^{\langle\mu}\omega^{\nu\rangle\gamma}\theta.
\end{align}
For conformal systems the complete set of possible third order terms was given in \cite{Grozdanov:2015kqa}. The specific form (\ref{TOSHEAR}) implements transport coefficients obtained from the RTA Boltzmann equation \cite{Jaiswal:2013vta}. We will here refer to Eq.~(\ref{TOSHEAR}) as the ``third-order" theory. Let us now simplify all of these equations for Bjorken flow. 


\vspace*{-2mm}
\section{Bjorken Flow}
\label{bjorken}
\vspace*{-2mm}

Milne coordinates $x^\mu=(\tau,x,y,\eta_s)$ (with $\tau=\sqrt{t^2{-}z^2}$ and $\eta_s=\tanh^{-1}(z/t)$) are the natural choice for describing ultra-relativistic heavy-ion collisions where the colliding nuclei approach each other approximately following light-cone trajectories. For Bjorken flow \cite{Bjorken:1982qr} of transversally homogeneous and longitudinally boost-invariant systems, macroscopic fields such as the energy density, pressure and shear stress, can depend neither on the transverse coordinates $(x,y)$ nor on the space-time rapidity $\eta_s$, but only on the longitudinal proper time $\tau$. The hydrodynamic evolution equations thus reduce to a set of coupled ordinary differential equations (ODEs) in $\tau$. The flow is irrotational ($\omega^{\mu\nu} = 0$) and unaccelerated ($u^\mu =(1,0,0,0)$, $\dot u^\mu = 0$), but (owing to the curvilinear nature of Milne coordinates) it has a non-zero local expansion rate, $\theta=1/\tau$, and  velocity shear, e.g. $\sigma^{\eta\eta} = -2/(3\tau^3)$.\footnote{%
	To avoid clutter we drop the subscript on $\eta_s$ whenever we use it as a sub- or superscript.}
Symmetries further constrain the shear tensor to be diagonal and space-like in Milne coordinates, leaving only one independent component which we take to be the $\eta\eta$ component: $\pi^{xx}=\pi^{yy}= -\tau^2 \pi^{\eta\eta}/2 \equiv \pi/2$.\footnote{%
	From here on we will use $\pi$ (not to be confused with the mathematical constant denoted by the 
	same symbol), or its normalized version $\bar\pi\equiv\pi/(\epsilon{+}P)=3\pi/(4\epsilon)$,
	as the independent shear stress component. Note that for Bjorken flow the Navier-Stokes value
	for $\pi$ is positive, $\pi_{_\mathrm{NS}}\geq0$.}
Using the following relations that hold for Bjorken flow,
\begin{align}
  &\dot\pi^{\langle\eta\eta\rangle} \!=\! -\frac{1}{\tau^2}\frac{d\pi}{d\tau},\quad
  \pi^{\langle\eta}_{\gamma}\sigma^{\eta\rangle\gamma} \!=\! -\frac{\pi}{3\tau^3},\quad
  \pi^{\langle\eta}_{\gamma}\pi^{\eta\rangle\gamma} \!=\! -\frac{\pi^2}{2\tau^2}, 
\nonumber\\ 
  &\pi^{\rho\langle\eta}\pi^{\eta\rangle\gamma}\sigma_{\rho\gamma} = -\frac{\pi^2}{2\tau^3}, \quad
  \nabla^{\langle\eta}\nabla_{\gamma}\pi^{\eta\rangle\gamma} = \frac{2\pi}{3\tau^4}, 
\label{identity}\\\nonumber
  &\nabla_{\gamma}\nabla^{\langle\eta}\pi^{\eta\rangle\gamma} = \frac{4\pi}{3\tau^4}, \quad
  \nabla^2\pi^{\langle\eta\eta\rangle} = \frac{4\pi}{3\tau^4}, \quad
  \pi^{\rho\gamma}\sigma_{\rho\gamma}=\frac{\pi}{\tau}, 
\end{align} 
the shear evolution equations \eqref{MIS}-\eqref{TOSHEAR} can be brought into the following generic form:
\begin{align}
  \frac{d\epsilon}{d\tau} &= -\frac{1}{\tau}\left(\frac{4}{3}\epsilon -\pi\right), 
\label{bde1}\\
  \frac{d\pi}{d\tau} &= - \frac{\pi}{\tau_\pi} + \frac{1}{\tau}\left[\frac{4}{3}\beta_\pi 
  - \left( \lambda + \frac{4}{3} \right) \pi - \chi\frac{\pi^2}{\beta_\pi}\right]. 
  \label{bde2}
\end{align}
The coefficients $\beta_\pi$, $a$, $\lambda$, $\chi$, and $\gamma$ appearing in Eq.~(\ref{bde2}) above and in Eq.~(\ref{dpibar}) below are tabulated in Table~\ref{coeff} for the three theories studied in this work.

\begin{table}[b!]
 \begin{center}
  \begin{tabular}{|c|c|c|c|c|c|}
   \hline
   & $\beta_\pi$ & $a$ & $\lambda$ & $\chi$ & $\gamma$ \\
   \hline
   MIS & $4P/5$ & 4/15 & 0 & 0 & 4/3 \\
   \hline
   DNMR & $4P/5$ & 4/15 & 10/21 & 0 & 4/3\\
   \hline
   Third-order & $\,4P/5\,$ & \,4/15\, & \,10/21\, & \,72/245\, & \,412/147\, \\
   \hline
  \end{tabular}
  \caption{Coefficients for the causal viscous hydrodynamic evolution of the shear stress in Bjorken 
  	flow for the three theories studied in this work.}
  \label{coeff}
 \end{center}
\end{table}

Since $\beta_\pi=4P/5=4\epsilon/15$, Eqs.~(\ref{bde1}) and (\ref{bde2}) are mutually coupled. Eq.~(\ref{bde2}) for the shear stress can be completely decoupled from the evolution of the energy density by rewriting it in terms of the normalized shear stress (inverse Reynolds number) $\bar\pi=\pi/(\epsilon{+}P)=\pi/(4P)$. Introducing at the same time the rescaled time variable \cite{Heller:2016rtz} $\bar{\tau}\equiv\tau/\tau_\pi$ (which is the inverse Knudsen number for Bjorken flow), Eq.~(\ref{bde1}) can be used to obtain the relation
\begin{equation}\label{vt1}
    \bar{\pi} = 3 \left( \frac{ \tau}{\bar{\tau}} \right) \frac{d \bar{\tau}}{d \tau} -2.
\end{equation}
Here we also used that for a conformal system $\epsilon{\,\propto\,}T^4$ and $T\tau_{\pi} = 5 \bar{\eta}=\mathrm{const.}$ where  $\bar{\eta} \equiv \eta/s$ is the specific shear viscosity. Equations (\ref{bde2}) and (\ref{vt1}) can now be combined to obtain a first-order nonlinear ordinary differential equation (ODE) for the inverse Reynolds number that is completely decoupled\footnote{\label{fn4}%
	This decoupling works as long as the relaxation time has a power law temperature dependence, 
	$\tau_\pi \propto T^{-\Delta}$, in which case in Eq.~(\ref{dpibar}) the coefficient multiplying 
	$d\bar{\pi}/d\bar{\tau}$ must be generalized to $\displaystyle{\frac{\Delta(\bar\pi{-}1){+}3}{3}}$.}
from the evolution of the energy density:\footnote{%
	Even if the relaxation time $\tau_\pi$ depends on temperature (as it does, for example, 
	in systems with conformal symmetry), the temperature evolution has been completely 
	absorbed into the rescaled time variable (inverse Knudsen number) $\bar\tau$.}
\begin{equation}
\label{dpibar}
   \left( \frac{ \bar{\pi} + 2 }{3} \right) \frac{d \bar{\pi}}{d \bar{\tau}} 
   = - \bar{\pi} + \frac{1}{\bar{\tau}} \left( a - \lambda \, \bar{\pi} - \gamma \, \bar{\pi}^2 \right).
\end{equation}
The three hydrodynamic theories studied here can be selected by choosing for $\lambda$ and $\gamma$ the appropriate combinations of constants given in Table~\ref{coeff}. All three theories share the same constant $a=4/15$, but we can solve Eq.~(\ref{dpibar}) numerically\footnote{%
	Equation (\ref{dpibar}) has the form of an Abel differential equation of the second kind for which, to the 
	best of our knowledge, an analytical solution does not exist.}
(and the very closely related Eq.~(\ref{rbde2}) to be discussed in Sec.~\ref{analy_sol} even analytically) for general $a$ and will therefore keep it as a free parameter until the end.

An important feature that we observe in Eq.~(\ref{dpibar}) is that the derivative $d\bar{\pi}/d\bar{\tau}$ diverges for $\bar{\pi}\to -2$, indicating a discontinuity in $\bar{\pi}$ at $-2$. This feature is not present in Eq.~(\ref{bde2}) and is merely an artifact of changing the evolution parameter from $\tau$ to $\bar{\tau}$. Moreover, we notice that the transverse pressure, $P_T\equiv P+\pi/2 =P(1+2\bar{\pi})$, becomes negative for $\bar{\pi}<-1/2$ indicating cavitation in the transverse direction which may lead to mechanical instability. Therefore one may conclude that $\bar{\pi}=-2$ already lies in the physically excluded region. In the next section, we study several other interesting properties inherent in the solutions of Eq.~(\ref{dpibar}).


\vspace*{-2mm}
\section{Gradient expansion, Lyapunov exponents and attractors}
\label{Lyapunov}
\vspace*{-2mm}

Consistent formulations of relativistic dissipative hydrodynamics involve short-lived non-hydrodynamic modes. These cannot be captured by a standard gradient expansion in terms of the Bjorken expansion rate $\theta=1/\tau$, causing such an expansion to be asymptotic, with zero radius of convergence \cite{Heller:2013fn}. Borel resummation of this divergent series leads to a hydrodynamic attractor which is well defined even for large gradients. The existence of an attractor in a theory is indicated by negative Lyapunov exponents\footnote{%
	Lyapunov exponents characterize the rate of separation of initially infinitesimally close trajectories in a dynamical system. A Lyapunov exponent is defined as $$|s(t)| \approx e^{\Lambda t} \, |s_0|,$$ where $s_0$ and $s(t)$ are the separation between two trajectories at initial time $t_0$ and at a	later time $t$, respectively. Negative $\Lambda$ indicates the rate at which the system approaches a 
	regular attractor.}
which govern the rate at which the system loses information about its initial conditions and evolves towards the attractor. In this section, we study the gradient expansion, Lyapunov exponents and their implications for attractor behavior in solutions of hydrodynamic equations for Bjorken flow. 

\vspace*{-2mm}
\subsection{Gradient expansion}
\label{sec4A}
\vspace*{-2mm}

To start with we consider the late proper-time expansion of $\bar{\pi}$. Substituting a power series ansatz for $\bar{\pi}$ in terms of powers of $1/\bar{\tau}$,
\begin{equation}
\label{series}
  \bar{\pi}(\bar{\tau}) = \sum_{n=0}^{\infty} \frac{c_n}{\bar{\tau}^n},
\end{equation}
into the non-linear ODE (\ref{dpibar}), one finds a recursion relation for the coefficients $c_n$ ($n\geq 1$) 
\begin{align}\label{coeff_eqn}
  c_n =  a \, \delta_{n,1} 
  &+ \left[ \frac{2}{3} \left( n-1 \right) - \lambda \right]  c_{n-1}  
\nonumber\\
  &+ \sum _{m=1}^{n} \Bigl(\frac{m{-}1}{3} - \gamma \Bigr)\, c_{n-m} \, c_{m-1},
\end{align}
with initial value $c_0{\,=\,}0$. The first non-vanishing coefficient $c_1{\,=\,}a$ corresponds to the Navier-Stokes term in the gradient expansion. 

 \begin{figure}[t]
 \begin{center}
  \includegraphics[width=\linewidth]{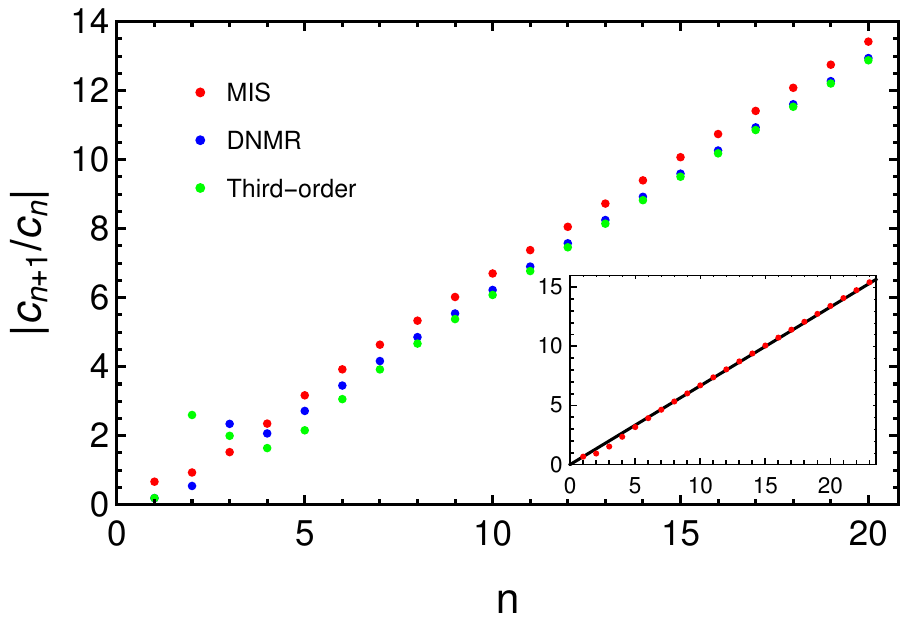}
 \end{center}
 \vspace{-0.8cm}
 \caption{Factorial behavior of coefficients (\ref{coeff_eqn}) for the three hydrodynamic theories indicated in the legend. In the inset the black line represents $|c_{n+1}/c_n|=\frac{2}{3}n$ and the red dots are for MIS theory.}
 \label{coeff_all}
\end{figure}

For large $n$ the behavior of the coefficients $c_n$ is dominated by the term proportional to $c_{n-1}$ in Eq.~(\ref{coeff_eqn}) and shows factorial growth. This is shown in Fig.~\ref{coeff_all} where the ratio of consecutive coefficients, $|c_{n+1}/c_n|$, is plotted against $n$; as shown in the inset this ratio is proportional to $n$ for $n\gtrsim 5$. Divergence of the hydrodynamic gradient expansion was found before in a variety of hydrodynamic theories \cite{Heller:2015dha, Florkowski:2016zsi, Denicol:2017lxn}. A new observation from Fig.~\ref{coeff_all} is that, while the coefficients of the DNMR and third-order theories are very similar, those for MIS theory feature a constant upward shift which can be attributed to $\lambda=0$ in Eq.~(\ref{coeff_eqn}). The inset of Fig.~\ref{coeff_all} shows the ratio $|c_{n+1}/c_n|$ for MIS theory ($\lambda{\,=\,}0$, red dots) compared with the same ratio obtained from Eq.~(\ref{coeff_eqn}) without the nonlinear last term (black line). One concludes that this nonlinear term plays no role in the asymptotic factorial growth of the expansion coefficients. The series solution (\ref{series}) is thus dominated by the term proportional to $c_{n-1}$ in Eq.~(\ref{coeff_eqn}). We verified that the same statement holds for the DNMR and third-order theories.

\vspace*{-2mm}
\subsection{``Effective MIS'' and Lyapunov exponents}
\label{sec4B}
\vspace*{-2mm}

\begin{figure}[t]
 \begin{center}
  \includegraphics[width=\linewidth]{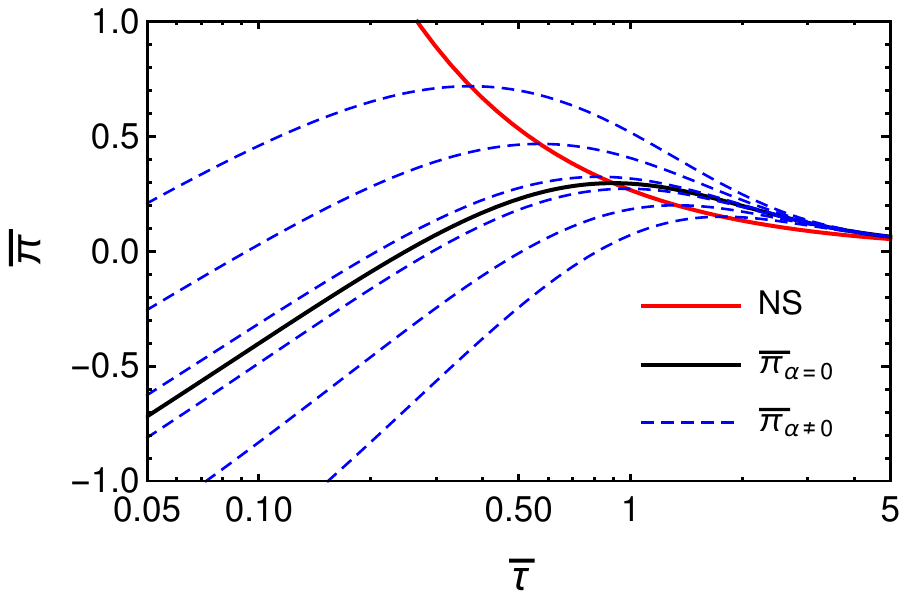}
 \end{center}
 \vspace{-0.8cm}
 \caption{Comparison of Eq.~(\ref{MIS_indefinite}) (solid black ($\alpha{\,=\,}0$) and dashed blue ($\alpha{\,\ne\,}0$) lines) with the asymptotic Navier-Stokes behavior (solid red line). See text for discussion.}
 \label{eff_MIS}
\end{figure}

The nonlinear last term in Eq.~(\ref{coeff_eqn}) arises from the quadratic terms $\propto\bar\pi^2,  d\bar\pi^2/d\bar\tau$ in Eq.~(\ref{dpibar}). At late times $\bar\pi{\,\ll\,}1$, hence $\bar\pi^2{\,\ll\,}\bar\pi$, 
and the nonlinear terms can be ignored  in (\ref{coeff_eqn}). The late-time behavior in MIS theory is thus controlled by 
\begin{equation}
\label{MIS_approx}
   \frac{2}{3}\dfrac{d\bar{\pi}}{d \bar{\tau} } = - \bar{\pi} + \frac{a}{\bar{\tau}}
\end{equation}
which we call ``effective MIS theory". Its analytic solution is 
\begin{equation} 
\label{MIS_indefinite}
   \bar{\pi} = \alpha\, e^{-\frac{3}{2}\bar{\tau}} 
                 + \frac{3a}{2}\, e^{-\frac{3}{2}\bar{\tau}}\, \mathrm{Ei} \left[\frac{3 \bar{\tau}}{2} \right]
\end{equation}
where $\mathrm{Ei}[z] = -\int_{-z}^{\infty} \left( e^{-t}/t \right) dt$ is the exponential integral and $\alpha$ is the integration constant. Equation (\ref{MIS_indefinite}) implies that the separation between two solutions for $\bar{\pi}$ that are initialized with different initial conditions is damped exponentially: 
\begin{equation}
\label{MIS_Lyapunov}
  \dfrac{\partial \bar{\pi}}{\partial\alpha} \sim \exp\!\left(-\frac{3}{2}\bar{\tau}\right).
\end{equation}
The negative Lyapunov exponent $\Lambda=-3/2$ in this equation confirms the existence of an attractor. This attractor, given by Eq.~(\ref{MIS_indefinite}) with $\alpha{\,=\,}0$, is shown in Fig.~\ref{eff_MIS} as the solid black line, together with a swarm of particular solutions of Eq.~(\ref{MIS_approx}) with different integration constants $\alpha{\,\ne\,}0$ (dashed lines) and the asymptotic Navier-Stokes solution (red solid line) for comparison. One sees that the attractor, as well as the other solutions with different initial conditions, join the asymptotic Navier-Stokes behavior after $\bar\tau{\,\gtrsim\,}3$, i.e. for Knudsen number Kn${\,\lesssim\,}0.3$. 

The solution (\ref{MIS_indefinite}) can also be written as
\begin{align}
\label{MIS_sol}
	\bar{\pi}(\bar\tau) = \bar\pi_0\, e^{-\frac{3}{2}\left(\bar{\tau}-\bar{\tau}_0\right)} 
	              + \frac{3a}{2}\, e^{-\frac{3}{2}\bar{\tau}} \int_{\bar{\tau}_0}^{\bar{\tau}}
	              \frac{e^{\frac{3}{2}\bar{\tau}'}}{\bar{\tau}'} d\bar{\tau}' 
\end{align}
where $\bar\pi_0\equiv\bar{\pi}(\bar{\tau}_0)$ is the initial condition at time $\bar{\tau}_0$. From Eq.~(\ref{MIS_sol}) one can extract the gradient expansion for $\bar{\pi}$ by integrating the last term by parts,
\begin{align}
   \int_{\bar{\tau}_0}^{\bar{\tau}} \frac{e^{\frac{3}{2}\bar{\tau}'}}{\bar{\tau}'} d\bar{\tau}' 
   &= \left[\sum_{n=1}^{m} (n{-}1)! \frac{e^{\frac{3}{2}\bar{\tau}'}}{(3\bar{\tau}'/2)^n}  
   \right]_{\bar{\tau}_0}^{\bar{\tau}} 
\nonumber \\
   & + \frac{3}{2} \, m! \, \int_{\bar{\tau}_0}^{\bar{\tau}} d\bar{\tau}' \, 
         \frac{e^{\frac{3}{2}\bar{\tau}'} }{(3\bar{\tau}'/2)^{m+1}},
\end{align}
resulting in
\begin{align}
   \bar{\pi} =  &e^{-\frac{3}{2}\left(\bar{\tau}-\bar{\tau}_0\right)} 
   \left( \bar\pi_0\, - a \sum_{n=1}^{m} \frac{\tilde{c}_n}{\bar{\tau}_0^n} \right) 
     + a\, \sum_{n=1}^{m} \frac{\tilde{c}_n}{\bar{\tau}^n} 
\\\nonumber 
   & + a\left(\frac{3}{2}\right)^2 m! \, e^{-3\bar{\tau}/2} \int_{\bar{\tau}_0}^{\bar{\tau}} d\bar{\tau}' \, 
         \frac{e^{\frac{3}{2}\bar{\tau}'} }{(3\bar{\tau}'/2)^{m+1}},
\end{align}
where $\tilde{c}_n= (n{-}1)!/(3/2)^{n{-}1}$. The first term within parentheses contains all dependence on the initial conditions memory of which is here seen to decay exponentially with a decay time of $\frac{2}{3}\tau_\pi$. The second term is the divergent gradient series (\ref{series}) up to order $m$. The third term can be thought of as the error in approximating the late time solution for $\bar{\pi}$ using a truncated gradient series of order $m$. Minimizing the error term with respect to $m$ would give an estimate of the optimal truncation for the gradient series at a given value of $\bar{\tau}$. 

We point out that Eq.~(\ref{MIS_sol}) also admits a convergent series in positive powers of $\bar{\tau}$ of the form
\begin{align}
	\bar{\pi} = 
	& \bar\pi_0\, e^{-\frac{3}{2}\left(\bar{\tau}-\bar{\tau}_0\right)} 
        + \frac{3a}{2}\, e^{-\frac{3}{2}\bar{\tau}} \, \log\left(\frac{\bar{\tau}}{{\bar{\tau}_0}}\right) 
\nonumber \\
	& + \frac{3a}{2}\, e^{-\frac{3}{2}\bar{\tau}} \sum_{n=1}^{\infty} \frac{(3/2)^n}{n! \, n} \, 
	       \left( \bar{\tau}^n - \bar{\tau}_0^n \right).
\end{align}	
It is obtained by expanding the exponential term $e^{\frac{3}{2}\bar{\tau}'}$ in the integral solution for $\bar{\pi}$ into a power series. This series is, however, of limited use at late times as it would require including a large number of terms for convergence. 

We conclude this subsection by offering another simple, yet interesting, form of the ``effective MIS'' solution for the inverse Reynolds number:
\begin{align}
	\bar{\pi} = \bar\pi_0\, e^{-\frac{3}{2}\left(\bar{\tau}-\bar{\tau}_0\right)} 
	              + \frac{3a}{2}\,  e^{-\frac{3}{2} \left[\bar{\tau} - \beta(\bar{\tau};\bar{\tau}_0)  \right]} \,
                          \frac{\bar{\tau}-\bar{\tau}_0}{\beta(\bar{\tau};\bar{\tau}_0)}.
\end{align}	
Here the function $\beta(\bar{\tau};\bar{\tau}_0)$ has units of time, and for any given pair $(\bar{\tau}_0, \bar{\tau})$ its value can be shown to lie in the interval $[\bar{\tau}_0,\bar{\tau}]$. To obtain this form we used the mean value theorem for the integral in Eq.~(\ref{MIS_sol}). It would be illuminating to derive an approximate analytical functional form for $\beta(\bar{\tau};\bar{\tau}_0)$; this is left for future work.

\subsection{Lyapunov exponents from linear perturbation}
\label{sec4C}

The exercise in the preceding subsection demonstrated exponential loss of memory of initial conditions in the ``modified MIS'' theory, on a time scale $\sim\frac{2}{3}\tau_\pi$. One arrives at similar conclusions without the assumption of ignoring in Eq.~(\ref{dpibar}) the terms quadratic in $\bar\pi$, by studying the fate of a linear perturbation around late-time solutions of the full equation (\ref{dpibar}). This is important because, as we will see further below, the effects on the evolution of $\bar\pi$ of the non-linear terms that were ignored in ``modified'' MIS theory last much longer than the memory-decay time over which initial-state information is erased.

We employ linear perturbation theory around the solution $\bar\pi$, i.e. we set $\bar{\pi}\to\bar{\pi} + \delta\bar{\pi}$ and insert this into the generic evolution equation (\ref{dpibar}). To simplify the nonlinear terms 
$\sim\bar\pi\delta\bar\pi$ we expand the solution $\bar\pi$ for late times according to Eq.~(\ref{series}), keeping only the first non-vanishing term which yields the Navier-Stokes approximation: $\bar{\pi}\approx a/\bar{\tau}$. Keeping terms up to first order in $1/\bar{\tau}$ we find
\begin{equation}\label{lin_pert_eqn}
  \dfrac{d(\delta\bar{\pi})}{d\bar{\tau}} = -\frac{3}{2}\delta\bar{\pi}\left[ 1+\frac{2\lambda-a}{2\bar{\tau}} \right],
\end{equation}
with the solution 
\begin{equation}\label{lin_pert_sol}
   \delta\bar{\pi} \sim \bar{\tau}^{\frac{3}{4}(a-2\lambda)} \exp\!\left(-\frac{3}{2}\bar{\tau}\right) 
   + \mathcal{O}\!\left(\frac{1}{\bar{\tau}^2}\right).
\end{equation}
The prefactor $\bar{\tau}^{\frac{3}{4}a}$ in front of the exponential decay factor is easily traced back to the quadratic term $\bar\pi\left(d\bar\pi/d\bar{\tau}\right)$ on the left hand side of Eq.~(\ref{dpibar}). The $\lambda$-dependence of this prefactor was not seen in the preceding subsection because $\lambda{\,=\,}0$ for MIS theory. The solution (\ref{lin_pert_sol}) exhibits the same negative Lyapunov exponent $\Lambda=-3/2$ as Eq.~(\ref{MIS_Lyapunov}). This shows that keeping the non-linear terms in Eq.~(\ref{dpibar}) does not change the initial state memory loss rate with which the system approaches the hydrodynamic attractor at late times.  This rate is entirely controlled by microscopic physics and independent of the precise macroscopic dynamical state of the expanding system.

\subsection{Lyapunov exponents from Borel resummation}
\label{sec4D}

The results in the preceding subsection may also be looked at from the perspective of Borel resummation, which replaces a divergent series $\sum_{n} a_n$ by \cite{Kleinert:2001ax}
\begin{equation}
   {\cal B}\left(\sum_{n} a_n \right) \equiv \int_{0}^{\infty}\, du \,\, e^{-u} \, \sum_{n} \frac{a_n}{n!} \, u^n.
\end{equation}	
Borel resummation interchanges the order in which the sum and integral in  $\sum_{n} a_n = \sum_{n} (a_n/n!) \int_0^{\infty} du \, e^{-u} \, u^n $ are performed. While for divergent series this yields inequivalent results, Borel resummation of asymptotic series can help with their interpretation, as shown below.  

Starting from the recursion relation (\ref{coeff_eqn}) and ignoring the non-linearities therein one finds the coefficients of the gradient series to be
\begin{equation}
    c_n = {\cal C} \,\, \frac{\Gamma(n{-}3\lambda/2)}{(3/2)^{n-3\lambda/2}}, \quad n\geq 1,
\end{equation}
where the normalisation ${\cal C}\equiv a(3/2)^{1{-}3\lambda/2}/\Gamma(1{-}3\lambda/2)$ ensures $c_1{\,=\,}a$. The Borel resummed version of Eq.~(\ref{series}) reads
\begin{align}
    {\cal B} \Bigg( \bar{\pi}({\bar{\tau}}) \Bigg) 
    &= {\cal C}\,\bar{\tau}^{-3\lambda/2} \int_{0}^{\infty} \frac{du}{u}\;e^{-(3/2)\bar{\tau}u}\, 
         u^{-3\lambda/2} \sum_{n=1}^{\infty} u^n  
\nonumber \\
    & = {\cal C}\,\bar{\tau}^{-3\lambda/2} \int_{0}^{\infty} du \, 
          e^{-(3/2)\bar{\tau}u}\,\frac{u^{-3\lambda/2}}{1-u}. 
\end{align}
The Borel integrand has a pole at $u{\,=\,}1$ which results in an ambiguous imaginary part of the sum, $\mathrm{Im}\left[ {\cal B}(\bar{\pi}) \right] = \pm \pi \,{\cal C} \,e^{-3\bar{\tau}/2} \bar{\tau}^{-3\lambda/2}$, arising from non-unique choices of deforming the contour around the pole. In order to remove the ambiguity in the Borel resummed gradient series one must include in Eq.~(\ref{series}) a ``non-hydrodynamic'' term $\delta \pi \sim e^{-3\bar{\tau}/2} \bar{\tau}^{-3\lambda/2}$.\footnote{%
	For a detailed study of this issue see the discussion of BRSSS theory \cite{Baier:2007ix} 
	in Ref.~\cite{Basar:2015ava}.}
Notice the similarity between this non-hydrodynamic term and that obtained via the method of linear perturbation in Eq.~(\ref{lin_pert_sol}).\footnote{%
	The extra factor of $\bar{\tau}^{3a/4}$ in the latter stems from non-linear terms in 
	Eq.~(\ref{coeff_eqn}) which were here ignored. It should be emphasized that when 
	considering the full non-linear version of Eq.~(\ref{dpibar}), not just one, but an 
	entire series of exponentially damped terms must be added to the gradient series, turning it
	into a trans-series \cite{Heller:2015dha, Basar:2015ava}. In this case the above-mentioned 
	non-hydrodynamic term plays the role of the leading order correction.}
%

\subsection{Numerical attractors for various theories}
\label{sec4E}

\begin{figure}[t]
 \begin{center}
  \includegraphics[width=\linewidth]{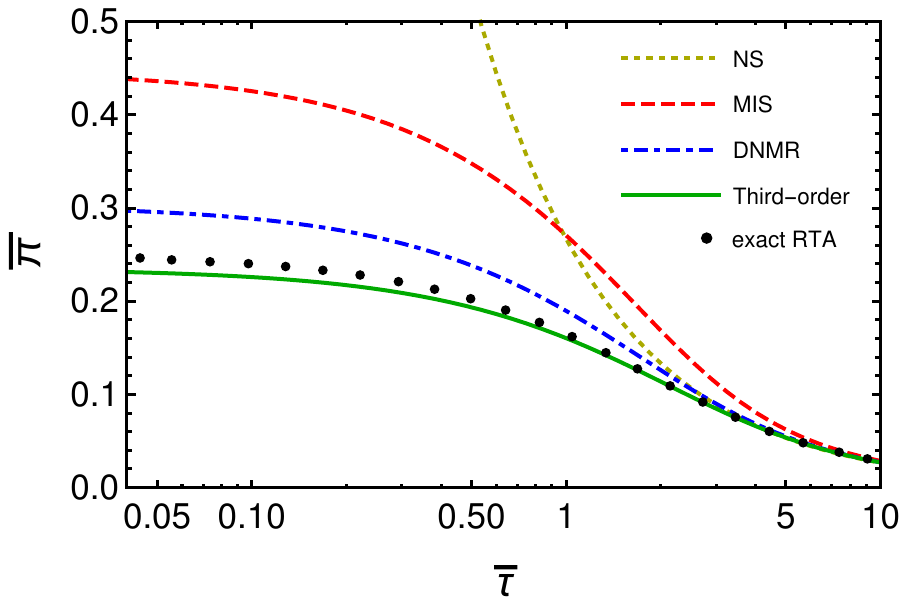}
 \end{center}
 \vspace{-0.8cm}
 \caption{Numerical attractors for the inverse Reynolds number $\bar\pi(\bar\tau)$ for the 
 	MIS (dashed line), DNMR (dash-dotted line), and third-order (solid line) theories, compared with the 
	exact numerical attractor of the RTA Boltzmann equation (filled circles). Also shown for 
	comparison is the Navier-Stokes solution (dotted).}
 \label{exact_attract}
\end{figure}

We close this section with a numerical study of the attractor solution of Eq.~(\ref{dpibar}) towards which specific solutions with generic initial conditions decay exponentially. To identify the attractor we follow the prescription outlined in Ref.~\cite{Heller:2015dha}: The initial condition for the attractor solution is obtained by imposing the boundary conditions that both $\bar{\pi}$ and $d\bar{\pi}/d\bar{\tau}$ remain finite as $\bar{\tau}\to 0$. As shown  in \cite{Heller:2015dha} this results in the quadratic equation $\gamma\,\bar{\pi}^2+\lambda\,\bar{\pi}-a=0$ for the initial value of $\bar{\pi}$. One finds two solutions, one of which (the positive root)  is stable and corresponds to the attractor solution whereas the negative root corresponds to a repulsor. With this initial condition we then solve Eq.~(\ref{dpibar}) numerically, for the different parameter combinations listed in Table~\ref{coeff}. The resulting attractors for the MIS, DNMR, and third-order theories are shown in Fig.~\ref{exact_attract} where we also compare them with the exact numerical attractor of the RTA Boltzmann equation \cite{Romatschke:2017vte}. Please note that, unlike any particular solution of Eq.~(\ref{dpibar}) which depends on both the initial condition $\bar\pi(\bar\tau_0)$ and the initial time $\bar\tau_0$ at which it is imposed, the attractors are universal, i.e. they attract any particular solution with initial conditions within their basin of attraction, irrespective of its starting time. 

In Fig.~\ref{exact_attract}, we show the attractor solutions for $\bar\pi$ for the MIS, DNMR, and third-order theories, as well as for the exact solution of RTA Boltzmann equation \cite{Strickland:2018ayk} and the Navier-Stokes solution. We see that of these the MIS attractor approaches the exact attractor most slowly,\footnote{%
	In light of footnote~\ref{fnMIS} this should perhaps not be too surprising.}
while the attractor of the third-order theory exhibits the best agreement with the exact RTA BE attractor, almost as good as the anisotropic hydrodynamic (aHydro) attractor studied in \cite{Strickland:2017kux, Chattopadhyay:2018apf}. This adds to the evidence of the superior performance of the third-order theory over different variants of second-order theories that are based on expansions around a locally isotropic momentum distribution.\footnote{%
	aHydro, a second-order approach that is based on an expansion around a self-consistently 
	adjusted {\it ellipsoidally deformed} local momentum distribution \cite{Florkowski:2013lya, 
	Tinti:2015xwa, Molnar:2016gwq, Strickland:2017kux, Martinez:2017ibh, McNelis:2018jho}, 
	performs even better than the third-order theory \cite{Chattopadhyay:2018apf}.}


\vspace*{-2mm}
\section{Approximate analytical solutions}
\label{analy_sol}
\vspace*{-2mm}

Up to this point we focused our discussion of the evolution of the inverse Reynolds number $\bar\pi$ on Eq.~(\ref{dpibar}) which holds for conformal systems where $T\tau_\pi{\,=\,}$const. It has the advantage of completely absorbing any dependence of the shear relaxation time $\tau_\pi$ on the energy density or temperature into the rescaled time variable (inverse Knudsen number) $\bar\tau$, but at the expense of not being able to solve this ODE analytically. In this section we derive analytical solutions for the evolution of $\bar\pi$ for Bjorken flow, at the expense of not being able to ensure the conformal relation $T\tau_\pi{\,=\,}$const. consistently with the evolution of the energy density. Instead, we find three separate classes of analytical solutions, corresponding to three different approximations of $\tau_\pi$ as a function of time. 

Starting from Eqs.~(\ref{bde1}),(\ref{bde2}), we decouple them as before by rewriting them in terms of the 
inverse Reynolds number $\bar\pi$ but without rescaling the time \cite{Chattopadhyay:2018pwe}:
\begin{align}
   &\frac{1}{\epsilon {\tau}^{4/3}} \frac{d(\epsilon {\tau}^{4/3})}{d \tau} = \frac{4}{3}\frac{\bar{\pi}}{\tau}, 
\label{rbde1} \\
   &\frac{d\bar{\pi}}{d \tau } =  - \frac{\bar{\pi}}{{\tau}_{\pi}} + \frac{1}{\tau} 
      \left( a - \lambda \bar{\pi} - \gamma {\bar{\pi}^2} \right). 
\label{rbde2} 
\end{align} 
In the following, we find analytical solutions of Eq.~(\ref{rbde2}), using different approximations for the form of shear relaxation time $\tau_\pi$. 

\vspace*{-2mm}
\subsection{Constant relaxation time}
\label{sec5A}
\vspace*{-2mm}

In this subsection we ignore the scaling of $\tau_\pi$ with temperature, by simply setting it constant. This constitutes a rather drastic violation of conformal symmetry by introducing, in addition to the inverse temperature $1/T$, a second, independent length scale $\tau_\pi$. In the following two subsections we will successively improve on this approximation.

The first analytical solution for the evolution in second-order hydrodynamics with Bjorken flow of the energy density and inverse Reynolds number for a constant shear relaxation time $\tau_\pi$ was found by Denicol and Noronha \cite{Denicol:2017lxn}. For completeness we briefly review this solution, generalizing it to the generic form (\ref{rbde1}),(\ref{rbde2}) of the evolution equations which also includes third-order hydrodynamics. Introducing again the rescaled time $\bar\tau=\tau/\tau_\pi$, for constant $\tau_{\pi}$ Eq.~(\ref{rbde2}) turns directly into
\begin{equation}\label{RBED3}
   \frac{d\bar{\pi}}{d \bar{\tau} } = 
   -\bar{\pi}  + \frac{1}{\bar{\tau}} \left( a - \lambda \bar{\pi} - \gamma {\bar{\pi}^2} \right)
\end{equation}
which is similar to Eq~(\ref{dpibar}) but without the nonlinearity on the left hand side.\footnote{%
	Equation (\ref{RBED3}) can be readily derived by setting $\Delta = 0$ in footnote~\ref{fn4}.}
As will be discussed below, this difference has important consequences for the attractor solutions and Lyapunov exponents.

Equation~(\ref{RBED3}) is a first-order nonlinear ODE of Riccati type which can be written as a second-order linear ODE with the help of the following transformation of variables,
\begin{equation}\label{VT} 
   \frac{1}{y}\frac{dy}{d\bar{\tau}}=\gamma\frac{\bar{\pi}}{\bar{\tau}} \quad 
   \Longleftrightarrow \quad \bar\pi = \frac{\bar{\tau}}{\gamma y}\frac{dy}{d \bar{\tau}},
\end{equation}
which turns Eq.~(\ref{RBED3}) into
\begin{equation} 
\label{DE}
    \frac{d^2 y}{d \bar{\tau}^2} + \left(1+ \frac{1+\lambda}{\bar{\tau}} \right) 
    \frac{dy}{d \bar{\tau}} - \frac{ a\gamma}{\bar{\tau}^2}\, y  = 0  .
\end{equation}
The general solution of this linear ODE can be expressed in terms of Whittaker functions $M_{k,m}(\bar{\tau})$ and $W_{k,m}(\bar{\tau})$ \cite{Denicol:2017lxn}\footnote{%
	Note that the authors of \cite{Denicol:2017lxn} used $\gamma=4/3$ and a different definition of $a$.}:
\begin{equation}\label{SOL2}
  y(\bar{\tau})=A  {\bar{\tau}}^k  e^{-\bar{\tau}/2}  \left[ M_{k,m}(\bar{\tau})+\alpha \  W_{k,m}(\bar{\tau}) \right].
\end{equation}
Here $k=-\frac{1}{2}(\lambda{+}1)$ and $m=\frac{1}{2}\sqrt{4a\gamma+\lambda^2}$ while $A$ and $\alpha$ are arbitrary constants. Substituting this solution in Eqs.~(\ref{VT}) and (\ref{rbde1}) one finds 
\begin{align}
   \bar{\pi}(\bar{\tau})=
   & \frac{(2 k{+}2 m{+}1) M_{k+1,m}(\bar{\tau}) - 2 \alpha W_{k+1,m}(\bar{\tau})}
             {2 \gamma \left[ M_{k,m}(\bar{\tau})+\alpha W_{k,m}(\bar{\tau}) \right]}, 
\label{const_pibar} \\ 
    \epsilon(\bar{\tau})=
    &\epsilon_0 \left(\frac{\bar\tau_{0}}{\bar\tau}\right)^{\!\frac{4}{3} \left(1-\frac{k}{\gamma}\right)}
       e^{-\frac{2}{3\gamma}\left( \bar{\tau}-{\bar{\tau}_0}\right)} 
\nonumber\\
     &\times \left( \frac{M_{k,m}(\bar{\tau}) + \alpha W_{k,m}(\bar{\tau})} 
                                 {M_{k,m}({\bar{\tau}}_0) + \alpha W_{k,m}({\bar{\tau}}_0 )}
                  \right)^{\frac{4}{3\gamma}} .
\label{const_eps}
\end{align} 
Here $\epsilon_0$ is the initial energy density at time ${\bar{\tau}}_0$, and the constant $\alpha$ encodes the initial normalized shear stress $\bar\pi_0$. Note that $\alpha$ can only take values for which the energy density is positive-definite for $\bar{\tau}>0$. 

It is easy to see from Eq.~(\ref{const_pibar}) that the solution for $\bar{\pi}(\bar{\tau})$ loses all memory about initial conditions at late times when $M_{k,m}(\bar{\tau})$ dominates over $W_{k,m}(\bar{\tau})$: for large arguments $\bar{\tau}\to\infty$ the ratio 
\begin{equation}
\label{asym}
   \frac{W_{k,m}(\bar{\tau})}{M_{k,m}(\bar{\tau})} \longrightarrow 
   \frac{\Gamma\left(m{-}k{+}\frac{1}{2}\right)}{\Gamma(2m{+}1)} \,\bar{\tau}^{2k}\, e^{-\bar{\tau}}
\end{equation}
decays exponentially. Thus at late times the terms proportional to $\alpha$ in (\ref{const_pibar}), (\ref{const_eps}) decay like $e^{-\bar\tau}$, corresponding to a Lyapunov exponent of $\Lambda{\,=\,}-1$. This obviously differs from $\Lambda{\,=\,}-\frac{3}{2}$ for the conformally invariant theories described by Eq.~(\ref{dpibar}); the difference is a direct consequence of the breaking of conformal symmetry by setting $\tau_\pi$ constant instead of $\propto 1/T$.

\vspace*{-2mm}
\subsection{Relaxation time from ideal hydrodynamics}
\label{sec5B}
\vspace*{-2mm}

A better approximation to Eq.~(\ref{dpibar}) can be obtained by setting $T\tau_\pi{\,=\,}$const. but, instead of using the exact time dependence of the temperature $T$, approximating the latter such that the evolution equations can still be integrated analytically. Seeing that for Bjorken flow the system asymptotically approaches local thermal equilibrium, we may approximate the time-dependence of $T$ at late times by the ideal fluid law \cite{Bjorken:1982qr}
\begin{equation}
\label{Bj_evol}
   T_{\rm id}(\tau) = T_0\left(\frac{\tau_0}{\tau}\right)^{1/3},
\end{equation}
where $T_0$ is the temperature at initial time $\tau_0$. For $T\tau_\pi=5\bar\eta$ this yields
\begin{equation}
\label{taupi_id}
   \tau_\pi(\tau) = b\,\tau^{1/3}, \quad\text{with}\quad b= \frac{5\bar\eta}{T_0\tau_0^{1/3}}.
\end{equation}
Using this to define the scaled time variable $\bar{\tau}\equiv\tau/\tau_\pi$ one finds 
$$\frac{d\bar{\tau}}{d\tau}=\frac{2}{3\tau_\pi} = \frac{2}{3b}\tau^{-1/3},$$ and Eq.~(\ref{rbde2}) turns into
\begin{equation}
\label{dpibar_id}
   \frac{2}{3}\frac{d\bar{\pi}}{d \bar{\tau} } =  -\bar{\pi}  
   + \frac{1}{\bar{\tau}} \left( a - \lambda \bar{\pi} - \gamma {\bar{\pi}^2} \right), 
\end{equation}
independent of $b$. This equation again misses the nonlinear term on the left hand side (l.h.s.) of Eq.~(\ref{dpibar}) and,  except for the factor $2/3$ on the l.h.s., has the same structure as Eq.~(\ref{RBED3}). Its analytical solution is therefore very similar to Eq.~(\ref{const_pibar}), except for a change of the argument of the Whittaker functions by a factor 2/3:
\begin{align}
   \bar{\pi}(\bar{\tau})=
   & \frac{(2 k{+}2 m{+}1) M_{k+1,m}(3\bar{\tau}/2) - 2 \alpha W_{k+1,m}(3\bar{\tau}/2)}
             {3 \gamma \ \left[ M_{k,m}(3\bar{\tau}/2)+\alpha W_{k,m}(3\bar{\tau}/2) \right]}, 
\label{id_pibar}
\\ 
   \epsilon(\bar{\tau})=
   & \epsilon_0 \left(\frac{\bar\tau_0}{\bar\tau}\right)^{\!\frac{4}{3} \left(\frac{3}{2}{-}\frac{k}{\gamma}\right)}\,
      e^{-\frac{1}{\gamma} (\bar{\tau}{-}{\bar{\tau}_0})} 
\nonumber\\       
    & \times 
    \left(\frac{M_{k,m}(3\bar{\tau}/2) + \alpha W_{k,m}(3\bar{\tau}/2)} 
                  {M_{k,m}({3\bar{\tau_0}/2}) + \alpha W_{k,m}(3\bar{\tau_0}/2)} 
    \right)^{\!\frac{4}{3 \gamma}} .
\label{id_eps}
\end{align} 
Here $k=-\frac{3 \lambda + 2}{4}$ and $m= \frac{3}{4} \sqrt{4 a \gamma +\lambda^2}$, different from 
Eqs.~(\ref{const_pibar})-(\ref{const_eps}). The asymptotic behavior (\ref{asym}) of the Whittaker functions now tells us that, for the choice (\ref{taupi_id}), memory of the initial conditions is lost exponentially according to $e^{-\frac{3}{2}\bar\tau}$, corresponding to the same Lyapunov exponent $\Lambda{\,=\,}-\frac{3}{2}$ as for the conformally invariant theories described by Eq.~(\ref{dpibar}).

\vspace*{-2mm}
\subsection{Relaxation time from Navier-Stokes evolution}
\label{sec5C}
\vspace*{-2mm}

We can further improve our approximation by accounting for first-order gradient effects in the evolution of the temperature, by replacing the ideal fluid law (\ref{Bj_evol}) by the Navier-Stokes result \cite{Kouno:1989ps, Muronga:2001zk, Baier:2006um}
\begin{equation}
\label{Bj_evol_NS}
    T_{_{\rm NS}} = T_0\left(\frac{\tau_0}{\tau}\right)^{1/3}
    \left[1 + \frac{2\bar{\eta}}{3\tau_0T_0}\left\{ 1- \left(\frac{\tau_0}{\tau}\right)^{2/3} \right\} \right].
\end{equation}
For $\bar{\eta}{\,=\,}0$ this reduces to (\ref{Bj_evol}). Substituting this into $T\tau_\pi=5\bar\eta$ we find
\begin{equation}
\label{taupi_NS}
    \tau_\pi = \frac{\tau^{1/3}}{d - \frac{2}{15}\tau^{-2/3}}\, , \qquad 
    d\equiv\left( \frac{T_0\tau_0}{5\bar{\eta}} + \frac{2}{15} \right)\tau_0^{-2/3}.
\end{equation}
For the scaled time variable $\bar{\tau}\equiv\tau/\tau_\pi$ we now have
\begin{equation}
\label{dtaubar_NS}
   \frac{d\bar{\tau}}{d\tau}=\frac{2}{3\tau_\pi}\left( 1 + \frac{2}{15\bar{\tau}} \right).
\end{equation}
Using this in Eq.~(\ref{rbde2}) one obtains
\begin{equation}
\label{dpibar_NS}
    \left( \frac{a/\bar{\tau}+2}{3} \right)\frac{d\bar{\pi}}{d \bar{\tau} } 
    =  -\bar{\pi}  + \frac{1}{\bar{\tau}} \left( a - \lambda \bar{\pi} - \gamma {\bar{\pi}^2} \right),
\end{equation}
independent of the constant $d$. This shares with Eq.~(\ref{dpibar_id}) the factor 2/3 on the l.h.s. which, as we saw in the preceding subsection, leads to the correct Lyapunov exponent for conformally symmetric systems. Comparing with Eq.~(\ref{dpibar}) one sees that they are identical up to the substitution $\bar\pi\mapsto a/\bar\tau$ (which is the first non-zero term in the series expansion (\ref{series})). This should not be surprising as the term within parenthesis on the l.h.s. of Eq.~(\ref{dpibar}) stems solely from the energy (or, equivalently, temperature) evolution equation (\ref{bde1}) which, when making the replacement $\bar\pi \mapsto a/\bar\tau$, leads to the Navier-Stokes solution $T_{_{\rm NS}}$. 

Comparing with the preceding subsection, this suggests that it might be possible to account for the time evolution of the temperature in the relation $\tau_\pi(\tau) \sim 1/T(\tau)$ with ever increasing precision by substituting the gradient series (\ref{series}) for $\bar\pi$ in the prefactor on the l.h.s. of Eq.~(\ref{dpibar}) and truncating it at increasingly higher order: zeroth order for ideal hydrodynamics, first order for Navier-Stokes dynamics, and so on. Unfortunately, this is not justified as Eq.~(\ref{series}) is an asymptotic (i.e. divergent) series. The consequences of this on the hydrodynamic attractor will be discussed in the next section.

Equation (\ref{dpibar_NS}) can be recast in a form similar to Eq.~(\ref{RBED3}) and solved again analytically: 
\begin{align}
   \bar{\pi}(\bar{\tau})=
   & \frac{(2 k{+}2 m{+}1) M_{k+1,m}(w)-2 \alpha W_{k+1,m}(w)}
              {3 \gamma \left[ M_{k,m}(w)+\alpha W_{k,m}(w) \right]}, 
\label{NS_pibar} \\ 
    \epsilon(\bar{\tau})=
    & \epsilon_0 \left(\frac{w_{0}}{w}\right)^{\!\frac{4}{3} \left(\frac{3}{2}-\frac{k}{\gamma}\right)}
       e^{-\frac{1}{\gamma} \left( \bar{\tau}-{\bar{\tau}_0} \right)}
\nonumber\\ 
     &\times \left(\frac{M_{k,m}(w) + \alpha W_{k,m}(w) } 
                                {M_{k,m}(w_0) + \alpha W_{k,m}(w_0)}
                  \right)^{\frac{4}{3 \gamma}} .
\label{NS_eps}
\end{align}
This looks formally identical to Eqs.~(\ref{id_pibar})-(\ref{id_eps}), except for the substitution $\frac{3}{2}\bar\tau\mapsto w\equiv \frac{3}{2}(\bar{\tau}+\frac{a}{2})$ in the arguments of the Whittaker functions on the right hand side (r.h.s.), together with modified definitions for the indices $k$ and $m$:
\begin{equation*}
   k=\frac{3a-4-6\lambda}{8}, \quad 
   m=\frac{3}{8}\sqrt{a^2 + 16a\gamma - 4a\lambda + 4\lambda^2}.
\end{equation*}
Using Eq.~(\ref{asym}) for the asymptotic behavior of the Whittaker functions we see that once again memory of the initial conditions is lost exponentially according to $e^{-\frac{3}{2}\bar\tau}$, with no effect from the constant shift of the time variable in the arguments of the Whittaker functions. This corresponds to the same Lyapunov exponent $\Lambda{\,=\,}-\frac{3}{2}$ as in the preceding subsection and, more generally, for all the conformally invariant theories described by Eq.~(\ref{dpibar}). 

We note in passing that including higher order terms $c_n/{\bar\tau}^n$ for $n{\,>\,}1$ in the gradient series while approximating $\bar\pi$ on the l.h.s. of Eq.~(\ref{dpibar_NS}) spoils its reducibility to the analytically solvable form explored in this work. Whether for such approximations Eq.~(\ref{dpibar}) can still be solved in terms of known functions remains to be seen.

We summarize this section by observing that, while the assumption of a fixed shear relaxation time $\tau_\pi$ (i.e. of a fixed temperature $T$ when writing $\tau_\pi\propto1/T$) leads to an incorrect Lyapunov exponent describing the rate of approach towards the hydrodynamic attractor, the correct decay rate is recovered as soon as one allows the temperature $T$ to vary with time hydrodynamically even if the exact time dependence is replaced by an approximation based on a truncated hydrodynamic gradient expansion.   


\vspace*{-2mm}
\section{Analytical attractors} 
\label{analy_attr}
\vspace*{-2mm}

\begin{figure}[t]
 \begin{center}
  \includegraphics[width=\linewidth]{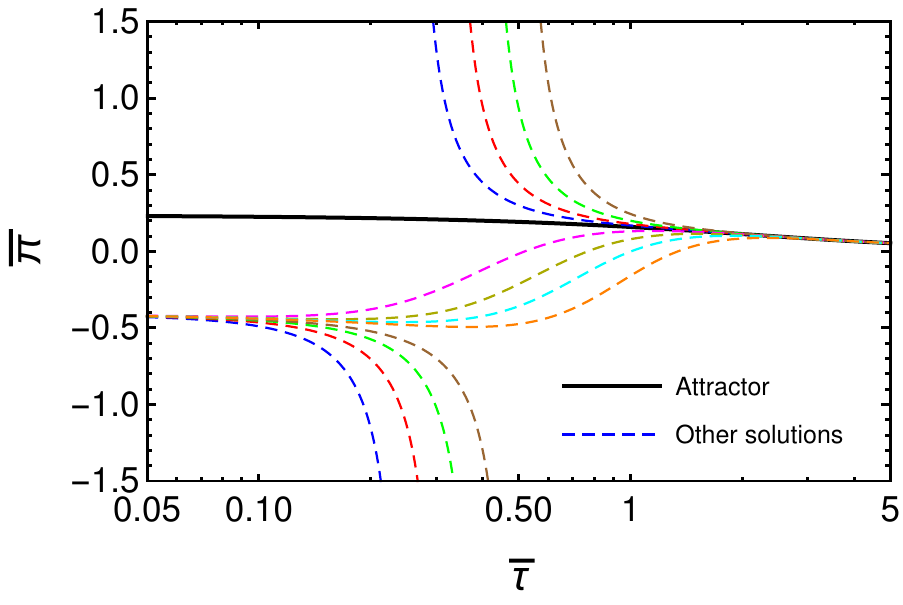}
 \end{center}
 \vspace{-0.8cm}
 \caption{Attractor behavior of the analytical solution (\ref{id_pibar}) for the third-order theory 
 	($\lambda{\,=\,}10/21$, $\gamma{\,=\,}412/147$). The two (attracting and repulsing) fixed 
	points at $\bar{\tau}\to0$ are clearly visible. Please note that some of the dashed lines have poles, 
	i.e. as $\bar{\tau}\to0$ they go to $+\infty$ before reappearing from $-\infty$ and approaching the 
	negative fixed point. 
 \label{att_bhvr}
 }
\end{figure}

In this section we investigate the hydrodynamic attractors associated with the analytic approximate solutions of the evolution equation for the inverse Reynolds number $\bar\pi$ found in the preceding section. In  Section~\ref{sec4E} we saw that, as $\bar\tau\to 0$, $\bar{\pi}$ can take one of only two finite values of opposite sign, and we identified the attractor as the unique solution which connects to the positive value. All other solutions were found to connect in the limit $\bar{\tau}\to 0$ to the negative value. An illustration of this generic behavior is shown in Fig.~\ref{att_bhvr} for the case of the attractor corresponding to the analytical solution (\ref{id_pibar}) for the third-order theory.

In this section we introduce the following procedure for identifying the hydrodynamic attractor  \cite{Chattopadhyay:2018pwe}: In terms of the parameter $\alpha$ encoding the initial condition for $\bar\pi$, we search for the value $\alpha_0$ at which the quantity
\begin{equation}
\label{psi}
   \psi(\alpha_0) \equiv \lim_{\bar{\tau} \to \bar{\tau}_0} 
   \frac{\partial\bar{\pi}}{\partial \alpha}\bigg{|}_{\alpha=\alpha_0}
\end{equation}
diverges at the time $\bar{\tau}_0$ where the two fixed points of the evolution trajectories (see Fig.~\ref{att_bhvr}) are located \cite{Behtash:2017wqg}. For Bjorken flow, this time is usually $\tau_0{\,=\,}0$. For the numerical solution of Eq.~(\ref{dpibar}) in Sec.~\ref{sec4E} this can be seen in Fig.~\ref{exact_attract}, and for the analytical solutions Eqs.~(\ref{const_pibar}) and (\ref{id_pibar}) in Secs.~\ref{sec5A} and \ref{sec5B}, respectively, this can be seen by studying their behavior near $\bar\tau=0$. The structural similarity of Eqs.~(\ref{NS_pibar}) and (\ref{id_pibar}) shows that for the solution (\ref{NS_pibar}) the fixed points are instead located at $w_0{\,=\,}0$, which corresponds to a negative (i.e. unphysical) longitudinal proper time $\bar{\tau}_0=-a/2$. This arises from the Navier-Stokes substitution $\bar\pi\mapsto a/\bar\tau$ on the l.h.s. of Eq.~(\ref{dpibar_NS}) which breaks down (and thus renders the analytic solution (\ref{NS_pibar}) unreliable) in the region $\bar\tau\ll1$.

We can use the same approach of studying the sensitivity to initial conditions to obtain the Lyapunov exponent $\Lambda$ from the formula
	\begin{equation}
	\label{Lambda}
	    \Lambda = \lim_{\bar{\tau}\to\infty}\,\dfrac{\partial}{\partial\bar{\tau}} 
	                      \left[ \ln\!\left( \dfrac{\partial\bar{\pi}}{\partial\alpha} \right)\right].
	\end{equation}
For the analytical solutions (\ref{const_pibar}), (\ref{id_pibar}), (\ref{NS_pibar}) this prescription reproduces the same results as obtained from the late-time behavior (\ref{asym}) of the Whittaker functions but its advantage is that it can also be used numerically where exact solutions for $\bar\pi$ are not available (such as for the numerical solutions of the generic equation (\ref{dpibar})). 

For the case shown in Fig.~\ref{att_bhvr} we have verified explicitly that indeed there are two fixed points at $\bar{\tau}_0=0$ and that only for the attractor solution, characterized by $\alpha_0{\,=\,}0$, $\psi(0)$ (defined in Eq.~(\ref{psi})) diverges. For all other solutions $\alpha\neq 0$ we found that $\psi(\alpha)=0$ in the limit $\bar{\tau}\to0$, indicating that they all converge to the negative branch, as shown in Fig.~\ref{att_bhvr}.

\begin{figure}[b]
 \begin{center}
  \includegraphics[width=\linewidth]{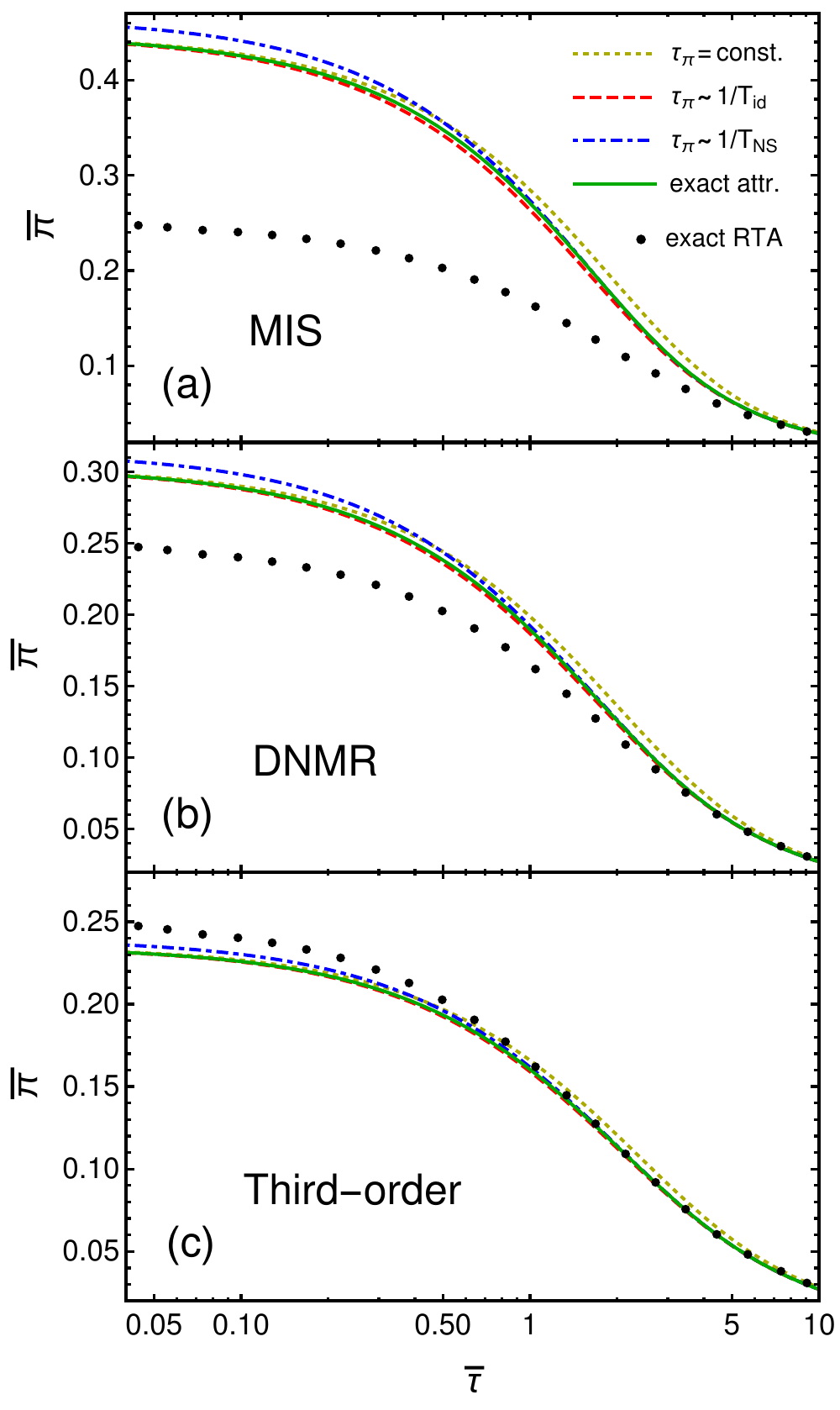}
 \end{center}
 \vspace{-0.8cm}
 \caption{Approximate analytical attractors for the MIS (a), DNMR (b), and third-order (c) theories,
 	compared with their exact numerical attractors (solid green lines) and the exact analytical 
	attractor for the RTA Boltzmann equation (black dots).}
 \label{att_all}
\end{figure}

The approximate analytical solutions (\ref{const_pibar}), (\ref{id_pibar}), (\ref{NS_pibar}) discussed in Sec.~\ref{analy_sol} can be written in the generic form 
\begin{align}
\label{generic}
   \bar{\pi}(w)=
   & \frac{(k{+}m{+}\frac{1}{2}) M_{k+1,m}(w) - \alpha W_{k+1,m}(w)}
             {\gamma |\Lambda| \left[ M_{k,m}(w)+\alpha W_{k,m}(w) \right]},
 \end{align}
with arguments and parameters for the three cases (i.e. for $\tau_\pi\propto 1/T$ with $T$ either constant or with time dependence taken from ideal or Navier-Stokes hydrodynamics) compiled for convenience in Table~\ref{T2}.
 %
\begin{table}[t]
 \begin{center}
  \begin{tabular}{|c|c|c|c|c|}
   \hline
   $T(\tau)$ & $w$ & $\Lambda$ & $k$ & $m$ \\
   \hline
   \phantom{$\bigg|$} const.  \phantom{$\bigg|$}
   & $\bar{\tau}$ & $\, -1 \ $ & $-\frac{1}{2} \left(\lambda{+}1\right)$ 
   & $\frac{1}{2} \sqrt{4a\gamma{+}\lambda^2}$  \\
   \hline
   \phantom{$\bigg|$} ideal  \phantom{$\bigg|$}
   & $\frac{3}{2}\bar{\tau}$ &\, $-\frac{3}{2}$ & $-\frac{3 \lambda +2}{4}$ 
   & $\frac{3}{4} \sqrt{ 4 a \gamma{+}\lambda^2  }$  \\
   \hline
   \phantom{$\bigg|$} NS  \phantom{$\bigg|$}
   & $\frac{3}{2} \left( \bar{\tau}{+}\frac{a}{2} \right)$ & $\,-\frac{3}{2}$ 
   & $-\frac{ 6\lambda + 4 - 3 a}{8}$ &\, $\frac{3}{8} \sqrt{16a\gamma{+}a^2{-}4a\lambda{+}4 \lambda^2}$\,  
   \\
   \hline
  \end{tabular}
  \label{power_coeff}
 \end{center}
\vspace*{-6mm}
\caption{Arguments and parameters of Eq.~(\ref{generic}) for the analytic approximations studied in 
	Secs.\,\ref{analy_sol}\,A, B, and C, respectively.
	\label{T2}}
\end{table}
 %
The attractor solutions are obtained from Eq.~(\ref{generic}) by setting the initial condition parameter $\alpha{\,=\,}0$:
\begin{align}
\label{gen_att}
   \bar{\pi}_\mathrm{attr}(w)=
   & \frac{k{+}m{+}\frac{1}{2}} {\gamma |\Lambda|}\, \frac{M_{k+1,m}(w)}{M_{k,m}(w)}.
 \end{align}
They are shown in Fig.~\ref{att_all} for the three different hydrodynamic theories discussed in this paper (MIS (a), DNMR (b), and third-order (c)) and compared with the corresponding exact numerical attractors as well as with the attractor for the exact analytical solution of the RTA Boltzmann equation \cite{Romatschke:2017vte} (the latter is, of course, the same in all three subpanels). Comparison of these attractors provides insights not only about the performance of the three different hydrodynamic theories as approximations to the underlying kinetic theory, but also about the relative accuracy of the additional approximations made in Sec.~\ref{analy_sol} in order to obtain analytical results.

For all three hydrodynamic theories, we note (especially at early times when the system is farthest away from local equilibrium) that the differences between the exact numerical attractors and their analytical approximations from Sec.~\ref{analy_sol} are significantly smaller than their discrepancy from the attractor of  the underlying kinetic theory. At late times the breaking of conformal symmetry by choosing a constant relaxation time leads generally to the largest difference between the exact numerical and analytically approximated hydrodynamic attractors; this may be attributed to the fact that the $\tau_\pi{\,=\,}$const. approximation underestimates the rate of approach towards the attractor by a factor 2/3. More surprisingly, the analytic approximation that performs best at late times (which uses $\tau_\pi(\tau) \sim 1/T_{_\mathrm{NS}}(\tau)$) performs worst at early times when compared with the exact numerical result. This reflects a different value for the fixed point of the inverse Reynolds number compared to the other analytic approximations and may be related to the fact that in this case the fixed point is shifted outside the physical region to $\bar\tau=-a/2$. As already seen in Fig.~\ref{exact_attract}, in comparison with the exact solution of the RTA Boltzmann equation, the third-order theory performs much better than both the MIS and DNMR theories, both when evaluated exactly numerically or approximately analytically. The only known theory that performs even better than the third-order theory studied here is second-order anisotropic hydrodynamics \cite{Bazow:2013ifa, Strickland:2017kux, Chattopadhyay:2018apf} which effectively resums terms of all orders in the inverse Reynolds number.

Numerical studies, such as those shown in Figs.~\ref{eff_MIS} and \ref{att_bhvr}, show that at late times $\bar\tau{\,>\,}1$ any initial deviation from the attractor approaches the attractor exponentially, with the Lyapunov exponents discussed before. The approximate analytical result (\ref{generic}) allows to understand this approach analytically over the entire range of $\bar\tau$, i.e. also for large Knudsen numbers $\bar\tau\ll 1$. Following the analysis \cite{Kurkela:2019set} we write\footnote{%
	Note that $\delta$ as defined in \cite{Kurkela:2019set} differs from ours by a factor 4/3.}
\begin{eqnarray}
\label{delta}
  \bar\pi(w) &=& \bar\pi_\mathrm{attr}(w) +\delta(w) 
\nonumber\\
                  &=& \left.\bar\pi(w)\right\vert_{\alpha=0} 
                     + \alpha \left.\frac{\partial\bar\pi}{\partial \alpha}(w)\right\vert_{\alpha=0}
                     + \mathcal{O}(\alpha^2)
\end{eqnarray}
where in the last expression we expanded the deviation $\delta$ to first order in the deviation of the initial value parameter $\alpha$ from the value $\alpha{\,=\,}0$ characterizing the attractor solution. The decay of the inverse Reynolds number $\bar\pi$ towards its attractor value is for small deviations\footnote{\label{delta_0}%
	It is worth pointing out that, for fixed initial deviation $\delta_0$ at initial time $\bar{\tau}_0$, the 
	corresponding initial state parameter $\alpha$ approaches 0 as $\bar{\tau}_0\to0$. Small $\delta_0$ 
	thus implies small $\alpha$ (especially for small values of $\bar\tau_0$), but not vice versa.}
$\delta$ given by
\begin{align}
\label{decay}
  \delta(w) = &-\frac{\alpha}{\gamma|\Lambda|} \frac{M_{k+1,m}(w)}{M_{k,m}(w)}
\\\nonumber
                    &\times\left[\left(k{+}m{+}\textstyle{\frac{1}{2}}\right) \frac{W_{k,m}(w)}{M_{k,m}(w)}
                                              + \frac{W_{k{+}1,m}(w)}{M_{k{+}1,m}(w)} \right].
\end{align}  
This is a function of the scaling variable $w\propto |\Lambda|\bar\tau$ which is proportional to the inverse Knudsen number $\bar\tau$. At first sight this suggests that the competition between the global expansion rate $1/\tau$ and the microscopic relaxation rate $1/\tau_\pi$ not only rules the evolution of the hydrodynamic attractor $\bar\pi_\mathrm{attr}$ itself, but also the way initial excursions of the inverse Reynolds number from this attractor decay as the full dynamical solution approaches the attractor. However, as first pointed out in \cite{Kurkela:2019set}, a deeper analysis exhibits that the dependence of the lifetime of such excursions on the interaction rate $1/\tau_\pi$ changes dramatically between early times $\tau{\,\ll\,}\tau_\pi/|\Lambda|$ and late times $\tau{\,\gg\,}\tau_\pi/|\Lambda|$.

\begin{figure}[h]
 \begin{center}
  \includegraphics[width=\linewidth]{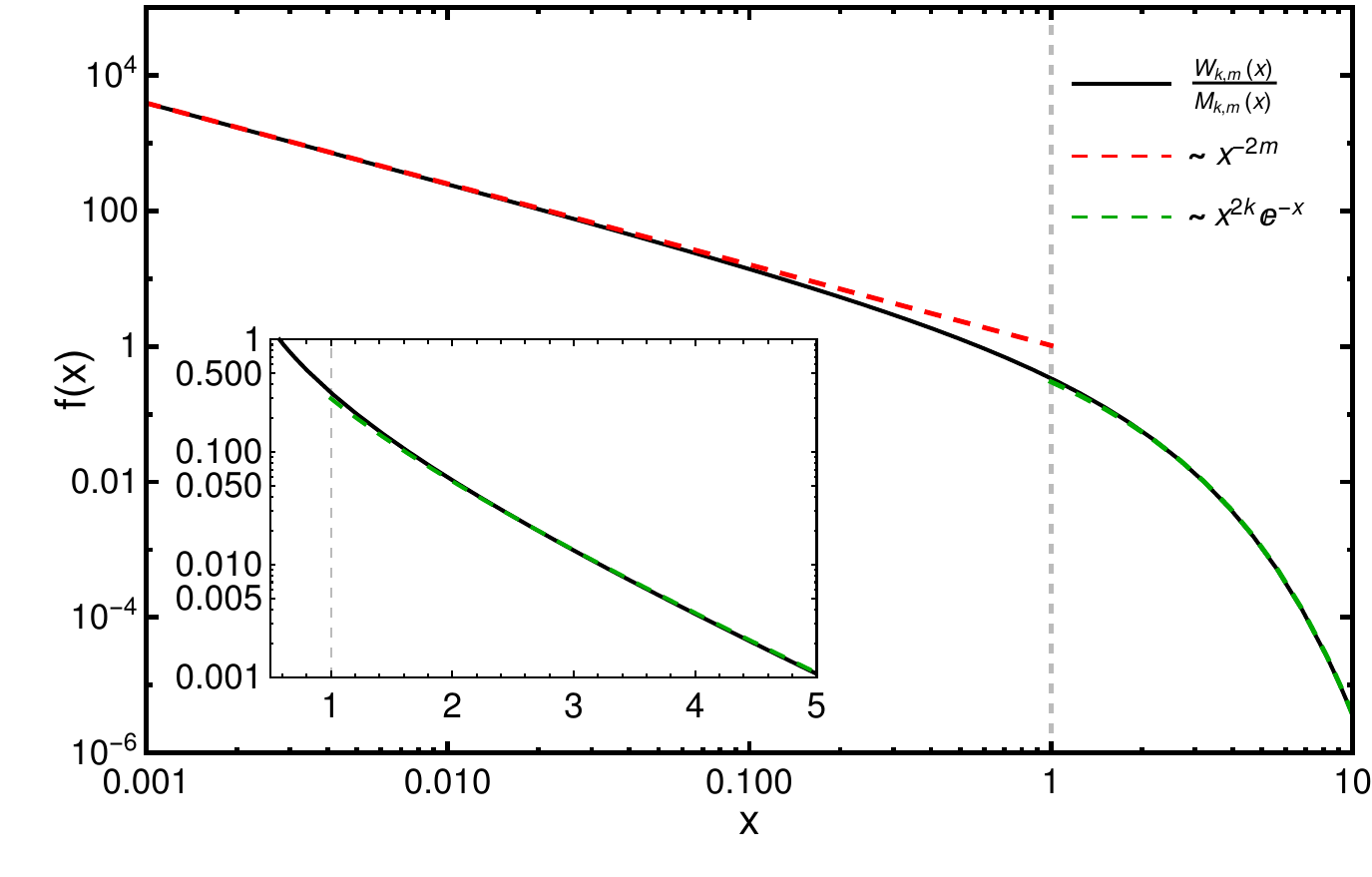}
 \end{center}
 \vspace{-0.4cm}
 \caption{The ratio $\frac{W_{k,m}(x)}{M_{k,m}(x)}$ as a function of the scaling variable $x$ 
 	(which is proportional to time in units of the microscopic relaxation time, i.e. to the inverse 
	Knudsen number). The exact result (solid black line) is compared with the first order 
	approximation (\ref{leading_order}) for small arguments $x<1$ (dashed red line) and the 
	analogous approximation (\ref{asym}) for large arguments $x>1$ (dashed green line).
	The power-law decay at early times manifests itself as an approximately straight line in the
	double-logarithmic representation of the main plot while the transition to exponential decay
	at late times leads to the approximately straight line behavior seen in the semi-logarithmic 
	inset plot. 
 	\label{F6}}
\end{figure}

Using the following Whittaker function properties for small arguments,
\begin{equation}
\label{leading_order}
  \left. \frac{W_{k,m}(x)}{M_{k,m}(x)} \right|_{x\to0} \!\!\!\!\approx   
  \frac{\Gamma[2m]}{\Gamma[m{-}k{+}\frac{1}{2}]} \, x^{-2m}, \quad
 \left. \frac{M_{k+1,m}(x)}{M_{k,m}(x)} \right|_{x\to0} \!\!\!\! \approx  1 ,
\end{equation}  
where the approximation holds at leading order, we see from Eq.~(\ref{decay}) that, in contrast to the exponential decay (\ref{asym}) at late times $\bar\tau\gg1$, the decay of excursions away from the attractor decay with a power law at early times $\bar\tau\ll 1$ \cite{Kurkela:2019set}. The transition from power law to exponential decay around $w\propto|\Lambda|\bar\tau=1$ is illustrated in Fig.~\ref{F6}.\footnote{%
	Note that the ``effective MIS'' approximation (\ref{MIS_indefinite}) discussed 
	in Sec.~\ref{sec4B} misses the early-time power-law decay of excursions away 
	from the hydrodynamic attractor. We found that early-time power-law decay of initial deviations from the attractor requires that in the generic evolution equation 
	(\ref{dpibar}) for the inverse Reynolds number at least one of the two coefficients $\lambda$ 	and $\gamma$ must be nonzero. This is not the case for the ``effective MIS" theory.}%
$^,$\footnote{\label{fn18}%
	A few comments regarding the domain of validity of the linearised approach are in order. Using Eq.~(\ref{generic}) to determine the deviation $\delta(w)$ from the attractor solution,	one readily checks that, for $w\ll1$, terms beyond the linear order in Eq.~(\ref{delta}) are 	proportional to $\alpha^{n} (w^{-2m})^n$ for $n\geq 2$. Even for fixed small values of 	$\alpha$, this series diverges for sufficiently early ``times'' $w^{2m}\ll\alpha$, where 	linearization in $\alpha$ must break down. This is closely related to footnote \ref{delta_0}: 	in order to keep $\delta_0 \equiv \delta(w_0)$ fixed for small values of $w_0$, $\alpha$ 	must be tuned down to ensure $\alpha w_0^{-2m}\ll1 $ such that higher-order contributions may be safely neglected. The breakdown of the linearised method at very early times owes 	itself to the presence of the repulsive fixed point at $w=0$, i.e., as long as $\alpha$ is not 	strictly equal to 0, all solutions, $\bar{\pi}(\alpha,w)$ will ultimately hit the repulsor at $w=0$, where the deviation from the attractor no longer stays infinitesimal. This mathematical feature of $\bar{\pi}(\alpha,w=0)$ being discontinuous at $\alpha=0$ was, in fact, used to uniquely determine the attractor solution via Eq.~(\ref{psi}) \cite{Chattopadhyay:2018pwe}.}

An immediate consequence of this discussion is that for early starting times $\tau_0{\,\ll\,}\tau_\pi/|\Lambda|$ (where, for sufficiently small $\delta$, 
$\delta(\bar\tau)\propto (|\Lambda|\bar\tau)^{-2m}$) the decay time $\tau_{1/e}$ (defined as the time over which the magnitude of $\delta$ decreases by a factor $1/e$) is given by $\tau_{1/e}=\rho\,\tau_0$ (where, to leading order in $\delta$ and $\bar\tau$, $\rho$ is a constant given in terms of the attractive fixed point $\bar\pi_{\mathrm{attr},0}$ by $\rho=(e^{1/2m}{-}1)^\frac{3}{\bar\pi_{\mathrm{attr},0}{+}2}$) and thus independent of $\tau_\pi$, whereas for late starting times $\tau_0{\,\gg\,}\tau_\pi/|\Lambda|$ it is given by $\tau_{1/e}=\tau_\pi/|\Lambda|$, independent of $\tau_0$. $\tau_{1/e}$ thus smoothly increases from 0 to $\tau_\pi/|\Lambda|$ (where it saturates) as $\tau_0$ increases from 0 to values much larger than $\bar\tau/|\Lambda|$. Sufficiently small excursions of the inverse Reynolds number from its attractor thus decay faster at early times $\tau{\,\ll\,}\tau_\pi/|\Lambda|$ than at late times $\tau{\,\gg\,}\tau_\pi/|\Lambda|$.  

Based on the observation that for $\tau\ll\tau_\pi/|\Lambda|$ the decay rate is controlled by the initial expansion rate $1/\tau_0$ and independent of the scattering rate $1/\tau_\pi$ whereas the opposite holds for  $\tau\gg\tau_\pi/|\Lambda|$, the authors of \cite{Kurkela:2019set} attribute the transition from power-law decay at early times to exponential decay at later times to a transition from the ``pre-hydrodynamic'' to the ``hydrodynamic'' stage, i.e. they see it as associated with the process of hydrodynamization. We have a different view of this matter: It is known from earlier numerical studies \cite{Strickland:2017kux, Behtash:2017wqg, Chattopadhyay:2018apf} of both Bjorken and Gubser flows that some hydrodynamic approximations (in particular anisotropic and third-order hydrodynamics, but to a somewhat lesser degree also DNMR theory) evolve the inverse Reynolds number very accurately (when compared with the evolution predicted by the underlying RTA Boltzmann equation) already at very early times $\bar\tau\ll 1$, even when initialized far away from the attractor. That is, not only do the attractors for these hydrodynamic approximations agree well with the exact attractor even for large Knudsen numbers, as shown for DNMR and third-order theories here in Fig.~\ref{exact_attract} and for anisotropic hydrodynamics in Fig.~3 of \cite{Strickland:2017kux}, but the theories also describe accurately {\em the evolution of the system towards the attractor} even when initialized far away from it. To us this implies that the system hydrodynamizes well before $\bar\tau=1/|\Lambda|$, and power-law rather than exponential decay of deviations from the attractor are not a tell-tale signature for ``pre-hydrodynamic'' behavior.


\vspace*{-2mm}
\section{Summary and conclusions}
\label{conclusion}
\vspace*{-2mm}

We studied analytically and numerically the evolution of the inverse Reynolds number in causal theories of second- and third-order relativistic viscous fluid dynamics for Bjorken flow. In this situation there is only a single non-vanishing component of the shear stress, describing an anisotropy between longitudinal and transverse pressure, which is generically very large at early times. For Bjorken flow the evolution of the associated inverse Reynolds number (i.e. the ratio of the shear stress to the enthalpy density of the system) decouples from that of the energy density and temperature and thus can be solved independently. When expressed as a function of time in units of the microscopic shear relaxation time (which measures the inverse Knudsen number of Bjorken flow), the solution is universal, i.e. independent of the specific shear viscosity of the medium. For three different macroscopic hydrodynamic theories, we studied these solutions, numerically and with various analytical approximations, their hydrodynamic attractors, the rate of initial state memory loss and approach to the attractor (expressed through Lyapunov exponents), and compared all these with the corresponding solution of the relativistic Boltzmann equation in RTA approximation which describes the underlying microscopic dynamics and can, for Bjorken flow, be solved exactly. 

When comparing the exact numerical solutions for the attractor of the inverse Reynolds number for the three different hydrodynamic theories studied here with the exact solution from the Boltzmann equation we find significant differences at early times, i.e. at large Knudsen numbers where the dynamics happens far away from equilibrium. While for the third-order theory the discrepancy remains always below 10\%,\footnote{%
	It has been shown elsewhere \cite{Strickland:2017kux, Chattopadhyay:2018apf} that the attractor for anisotropic hydrodynamics (aHydro) is even closer to the exact Boltzmann attractor than the one for the third-order theory.}
it increases to ${\sim\,}30\%$ for DNMR theory and to almost a factor of 2 for MIS theory (both being second-order hydrodynamic theories). Compared to these, the additional discrepancies caused by the various approximations we made to arrive at analytical solutions for the dynamics of $\bar\pi$ are small. At late times (small Knudsen numbers) all attractors approach the Navier-Stokes solution. Again third-order hydrodynamics is closest to the exact solution of the underlying Boltzmann kinetics, but the excellent agreement is spoiled somewhat if conformal symmetry is broken by an approximation that sets the relaxation time $\tau_\pi$ to a constant rather than allowing it to vary inversely with the temperature.  

As $\tau\to0$, the Navier-Stokes value of the inverse Reynolds number diverges while its attractor value approaches the finite value $\bar\pi=0.25$, corresponding to a shear stress over thermal pressure ratio $\pi/P{\,=\,}1$. The Lyapunov exponent associated with the evolution of $\bar\pi$, $\Lambda{\,=\,}-\frac{3}{2}$, indicates that even far-from-equilibrium initial conditions, $\pi_0/P_0\gg1$, relax exponentially to the attractor value with a decay time of order $\frac{2}{3}\tau_\pi=\frac{10}{3}\frac{\bar\eta}{T}$. For minimal specific shear viscosity $\bar\eta=\frac{1}{4\pi}$ and a medium temperature of, say, $T{\,=\,}0.5$\,GeV, this corresponds to a decay time of $\approx 0.1$\,fm/$c$. Subsequently, the hydrodynamic evolution follows essentially the hydrodynamic attractor of the theory which agrees, within the precision stated above, with the exact attractor associated with of the underlying microscopic Boltzmann kinetics. 

While the ODE describing the evolution of the inverse Reynolds number for Bjorken flow is easily solved on a computer, with arbitrary precision, the analytic approximations studied here are surprisingly accurate, and they yield valuable insights into the details of initial state memory loss and the approach to attractor dynamics in Bjorken flow. Similar methods may be applicable to different situations (for example, Gubser flow, which is physically quite different from Bjorken flow but shares with it many mathematical similarities) where they can lead to similarly valuable qualitative insights. Generalization to Gubser flow should be particularly interesting because it does not thermalize at late times but rather approaches an asymptotic free-streaming state. In such a situation the questions of initial state memory loss and the approach to a hydrodynamic attractor \cite{Behtash:2017wqg, Chattopadhyay:2018apf, Denicol:2018pak} have not yet been fully understood.

\begin{acknowledgments}

S.J. and A.J. acknowledge kind hospitality of IISER Pune where parts of this work were completed. We gratefully acknowledge interesting discussions with A.~Behtash, S. Kamata, M.~Martinez and H. Shi as well as the stimulating effect of Ref.~\cite{Kurkela:2019set} which appeared simultaneously when the present work was first circulated and which led to several additional insights reported in Sec.~\ref{analy_attr} of this final version of our work.  U.H. thanks Urs Wiedemann for an important clarifying discussion of Ref.~\cite{Kurkela:2019set}. A.J. is supported in part by the DST-INSPIRE faculty award from the Indian Department of Science and Technology under Grant No. DST/INSPIRE/04/2017/000038. The research of U.H. and C.C. was supported in part by the U.S. Department of Energy (DOE), Office of Science, Office for Nuclear Physics under Award No. \rm{DE-SC0004286} and in part by the National Science Foundation (NSF) within the framework of the JETSCAPE Collaboration under Award No. \rm{ACI-1550223}. S.J. thanks Pranav S. and C.C. thanks Yuri Kovchegov for illuminating discussions. U.H. acknowledges the kind hospitality of the Institut f\"ur Theoretische Physik at the J. W. Goethe-Universit\"at during a research visit supported by a Research Prize from the Alexander von Humboldt Foundation. He thanks  Micha{\l} Heller and Viktor Svensson for fruitful discussions that motivated some of the research reported here. 

\end{acknowledgments}

\bibliographystyle{JHEP}
\bibliography{reference}

\end{document}